\documentclass[pre,aps,preprint,showpacs]{revtex4}
\usepackage{graphicx}
\usepackage{amsfonts}
\usepackage{amssymb}
\usepackage{amsmath}

\begin{document}

\title{Organizing Principles for Dense Packings of Nonspherical Hard Particles:\\
Not All Shapes Are Created Equal}

\author{Salvatore Torquato}
\email{torquato@princeton.edu}
\affiliation{
Department of Chemistry, Department of Physics,
Princeton Center for Theoretical Science,
Program of Applied and Computational Mathematics,
Princeton Institute of the Science and Technology of Materials,
Princeton University, Princeton, New Jersey 08544, USA}
\author{Yang Jiao}
\email{yjiao@princeton.edu}
\affiliation{Princeton Institute of the Science and Technology of Materials,
Princeton University, Princeton, New Jersey 08544, USA}

\begin{abstract}

We have recently devised organizing principles to obtain maximally
dense packings of the Platonic and Archimedean solids, and certain
smoothly-shaped convex nonspherical particles [Torquato and Jiao,
Phys. Rev. E {\bf 81}, 041310 (2010)]. Here we generalize them in
order to guide one to ascertain the densest packings of other
convex nonspherical particles as well as {\it concave} shapes. Our
generalized organizing principles are explicitly stated as four
distinct propositions. These organizing
principles are applied to and tested against the most comprehensive set of both convex and
concave particle shapes examined to date, including Catalan solids, prisms,
antiprisms, cylinders, dimers of spheres and various concave
polyhedra. We demonstrate that all of the densest known packings
associated with this wide spectrum of nonspherical particles are
consistent with our propositions. Among other applications, our
general organizing principles enable us to construct analytically
the densest known packings of certain convex nonspherical
particles, including spherocylinders, ``lens-shaped'' particles,
square pyramids and rhombic pyramids. Moreover, we show how to
apply these principles to infer the high-density equilibrium
crystalline phases of hard convex and concave particles. We also
discuss the unique packing attributes of maximally random jammed
packings of nonspherical particles.


\end{abstract}


\maketitle


\section{Introduction}

Dense packings of hard particles have served as useful models to
understand the structure of low-temperature states of matter, such
as liquids, glasses, and crystals \cite{Be65, Zallen, ChLu00},
granular media \cite{Edwards}, heterogeneous materials
\cite{SalBook}, and biological systems (e.g., tissue structure,
cell membranes, and phyllotaxis) \cite{Li01, Pu03, Ge08, ToStRMP}.
Much of what we know about dense packings concerns particles of
spherical shape \cite{Conway, Henry, hales,Bezdek,Parisi,ToStRMP}, and hence it is
useful to summarize key aspects of the {\it geometric-structure}
classification of sphere packings \cite{ToStRMP} in order to place our results
for dense packings of nonspherical particles in their proper
context.

The geometric-structure classification naturally emphasizes that
there is a great diversity in the types of attainable jammed
sphere packings with varying magnitudes of overall order
(characterized by scalar order metric $\psi$ that lies in the interval $[0,1]$), packing fraction
$\phi$, and other intensive parameters (e.g, mean contact number) \cite{ToTrDe00, ToStRMP,
JiaoJAP11}. The notions of ``order maps'' \cite{ToTrDe00} in
combination with mathematically precise ``jamming categories''
\cite{ToSt01JPC} enable one to view and characterize well-known
packing states, such as the densest sphere packing (Kepler's
conjecture) \cite{hales} and maximally random jammed (MRJ) packings as extremal
states in the order map for a given jamming category
\cite{ToStRMP}. Indeed, this picture encompasses not only these
special jammed states, but an uncountably infinite number of other
packings, some of which have only recently been identified as
physically significant, e.g., the jamming-threshold states (least
dense jammed packings) \cite{SalTunnel} as well as extremal states
between these and the MRJ point \cite{JiaoJAP11}. Figure 1 shows a
schematic order map for frictionless spheres in the $\phi$-$\psi$
plane \cite{fn_ordermap} based on our latest knowledge
\cite{ToStRMP, JiaoJAP11}. 

\begin{figure}[htbp]
\begin{center}
$\begin{array}{c}
\includegraphics[width=8.0cm,keepaspectratio]{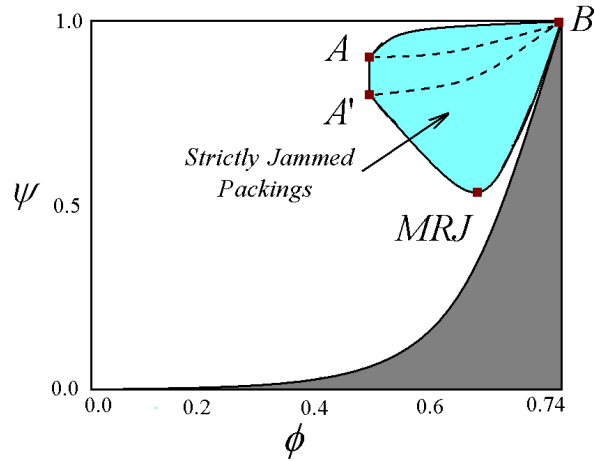}
\end{array}$
\end{center}
\label{fig_ordermap} \caption{(Color online) Schematic order map in the
density-order $\phi$-$\psi$ plane for {\it strictly} jammed (mechanically
stable) packings
\cite{ToSt01JPC} of frictionless identical spheres in three
dimensions (adapted from Ref.~\cite{ToStRMP}). White and blue (or light gray in print version)
regions contain the attainable packings, blue regions (or light gray in print version) represent
the jammed subspaces, and the dark gray region contains no packings.
The locus of points $A$-$A^\prime$ correspond to the
lowest-density jammed packings (thought to be ``tunneled crystals"
with contact number per particle of exactly 7 \cite{SalTunnel}).
The locus of points $B$-$B^\prime$ correspond to the densest
jammed packings with contact number per particle of exactly 12.
Point MRJ represents degenerate maximally random jammed states
with a mean contact number per particle $Z=6$, i.e., the most
disordered states subject to the jamming constraint. Although much
less is known about the corresponding order maps for nonspherical
particles, our current understanding of some of the extremal
points shows that they can be quite distinct from their sphere
counterparts, as discussed in Secs. III-V.}
\end{figure}

In recent years, scientific attention has broadened from the study
of dense packings of spheres (the simplest shape that does not
tile Euclidean space) \cite{hales} to dense packings of disordered
\cite{DoSiSaVa04, Ohern09, JiStTo10, DeCl10, tetrah1,
spherocylinder, particulogy, tetrah3, Paul10, Fisher10, Fisher11,
PlatonicMRJ, ChaseMRJ} and ordered \cite{Bezkuper, Wills, DonevEllip,
JiaoSuperdisk, JiaoSuperball, NiSuperball, Be00, Co06, ToJi09,
ToJi09b, tetrah2, Ka10, ToJi10, Ch10, kallus11, tiling,
TrunTetrah, GlozterTrunTetrah, Laura09, Dijkstra, Aga11, NatMat12}
nonspherical particles. The focus of the present paper is on the
densest packings of {\it congruent} nonspherical particles, both
convex and concave shapes. We will show that both the symmetry and
local principal curvatures of the particle surface play a crucial
role in how the rotational degrees of freedom couple with the
translational degrees of freedom to determine the maximal-density
configurations. Thus, different particle shapes will possess
maximal-density configurations with generally different packing
characteristics.  We have recently devised organizing principles 
(in the form of conjectures) to obtain maximally dense packings of a 
certain class of convex nonspherical hard particles \cite{ToJi09,ToJi09b,ToJi10}. Interestingly, 
our conjecture for certain centrally symmetric polyhedra have been 
confirmed experimentally very recently \cite{NatMat12}.
Here we generalize these organizing principles to other convex particles
as well as concave shapes \cite{SalAPSTalk}. The principles for concave shapes were
implicitly given in Ref.~\cite{JiaoSuperdisk, ToJi10} and were
explicitly stated (without elaboration) elsewhere
\cite{SalAPSTalk}.

Tunability capability via particle shape
provides a means to design novel crystal, liquid and glassy states
of matter that are richer than can be achieved with spheres (see
Fig. 2). While it is seen that hard spheres exhibit
entropically driven disorder-order transitions, metastable
disordered states, glassy jammed states, and ordered jammed states
with maximal density, the corresponding phase diagram for hard
nonspherical particles will generally be considerably richer in
complexity due to the rotational degrees of freedom and smoothness
of the particle surface (see Sec. IV for further details).

\begin{figure} [htbp]
\begin{center}
$\begin{array}{c}
\includegraphics[width=8.0cm,keepaspectratio]{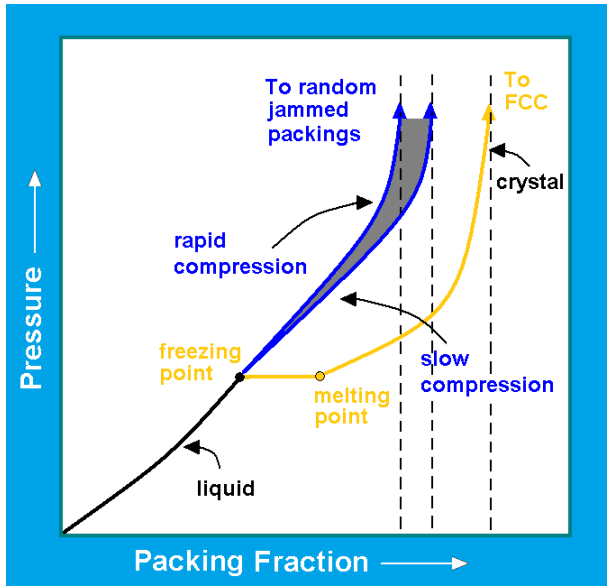}
\end{array}$
\end{center}
\label{fig_phasediagram} \caption{(Color online) The isothermal phase behavior of
three-dimensional hard-sphere model in the pressure-packing
fraction plane, adapted from Refs.~\cite{SalBook, ToStRMP,
rint96}. Three different isothermal densification paths by which a
hard-sphere liquid may jam are shown.  An infinitesimal
compression rate of the liquid traces out the thermodynamic
equilibrium path (shown in yellow or light gray in print version), including a discontinuity
resulting from the first-order freezing transition to a crystal
branch.  Rapid compressions of the liquid while suppressing some
degree of local order (curves shown in blue or dark gray in print version) 
can avoid crystal nucleation
(on short time scales) and produce a range of amorphous metastable
extensions of the liquid branch (shown in black) that jam only at the their
associated terminal densities. The corresponding phase diagrams
for nonspherical particles will generally be considerably more
complex (see Sec. IV for further details).}
\end{figure}


Although the focus of the present paper will be the development of
organizing principles to obtain dense packings of nonspherical
particles in three-dimensional Euclidean space $\mathbb{R}^3$, it
is useful to note that such principles have been put forth for
two-dimensional convex particles \cite{Toth, kuper2,pach}. In
particular, Fejes-T{\'o}th \cite{Toth} showed that the densest
packing of a centrally symmetric particle in two-dimensional
Euclidean space $\mathbb{R}^2$ can be obtained by circumscribing
the particle with the minimal centrally symmetric hexagon, which
tiles $\mathbb{R}^2$. This leads to a Bravais-lattice packing
(defined in Sec. II) of the centrally symmetric particle under
consideration. A key element in his argument is the fact that
local optimality is consistent with the global optimality in two
dimensions (e.g., a centrally symmetric hexagon can always
tessellate the space), which unfortunately generally does not hold
in three dimensions. For convex particles in $\mathbb{R}^2$
without central symmetry, Kuperberg and Kuperberg \cite{kuper2}
showed that dense packings of a specific particle can be obtained
by constructing a \textit{double-lattice} packing of the particle,
i.e., a packing composed of a Bravais-lattice packing of the
particle itself and a lattice packing of the center-inversion
image of the particle. They showed that the maximal-density
double-lattice packing of a specific shape can be obtained by
minimizing the area of certain parallelogram associated with the
packing \cite{kuper_bound}. We note that the double-lattice
packing amounts to a Bravais-lattice packing of a centrally
symmetric dimer (i.e., pair) of the original particles.

Generalizing the aforementioned organizing principles to
three-dimensional Euclidean space $\mathbb{R}^3$, which is
fundamentally different from $\mathbb{R}^2$, is highly nontrivial.
In $\mathbb{R}^2$, obtaining the maximally dense packing of a
specific particle shape can usually be reduced to a local problem,
because the densest local clusters of the particles are frequently
consistent with the globally densest packing. However, this
principle generally does not apply in $\mathbb{R}^3$, since the
densest local packing clusters are usually ``geometrically
frustrated'' \cite{hales} and, hence, inconsistent with the
globally densest packing. For example, the densest local packing
of congruent two-dimensional circles corresponds to an equilateral
triangle, which can tessellate $\mathbb{R}^2$. This enables one to
easily show that the maximally dense packing of circles in
$\mathbb{R}^2$ is the triangular-lattice packing \cite{Conway}. In
$\mathbb{R}^3$, the densest local packing of congruent spheres
corresponds to a regular tetrahedron, which cannot completely
fill space \cite{Co06, kallus_bound}. This was one reason why it
took nearly 400 years to prove that one of the maximally
dense packings of congruent spheres is achieved by the face-centered-cubic
lattice packing (Kepler's conjecture) and its stacking variants \cite{hales}. For nonspherical
particles in $\mathbb{R}^3$, the situation is even more complex \cite{Bezdek}
and any proofs establishing the densest packings of nonspherical particles 
will be considerably more challenging than for spheres.
For example, in the densest known packing of regular tetrahedra,
the centrally symmetric fundamental packing unit is composed of
four tetrahedra \cite{Ka10, ToJi10, Ch10}, instead of two as in a
double-lattice packing in $\mathbb{R}^2$.
It is therefore extremely useful if general organizing principles 
can be devised to lead one to obtain densest packings of nonspherical particles.

In this paper, we generalize the organizing principles devised for
obtaining the maximally dense packings of the Platonic and
Archimedean solids and certain smoothly-shaped convex particles
(e.g., superballs) to guide [41, 42, 45] one to find the densest packings of
other convex nonspherical particles as well as to concave shapes.
We explicitly state our generalized organizing principles as four
distinct propositions, which are applied and tested to the most
comprehensive set of both convex and concave particle shapes examined to date,
including Catalan solids, prisms, antiprisms, cylinders, dimers of
spheres and various concave polyhedra. All of the densest known
packings associated with the aforementioned large set of
nonspherical particles are consistent with our propositions.
Moreover, we show that how they can be applied to construct
analytically the densest known packings of certain convex
nonspherical particles, including spherocylinders and
``lens-shaped'' particles that are centrally symmetric, as well as
square pyramids and rhombic pyramids that lack central symmetry.
Although we do not provide rigorous mathematical proofs for our 
propositions here, we expect that they will either ultimately lead to strict proofs or at least 
provide crucial guidance in obtaining such proofs for some nonspherical
shapes.

Moreover, we show how to apply our organizing principles to infer the
high-density equilibrium crystalline phases of hard convex and
concave particles. The unique packing attributes of maximally
random jammed packings of nonspherical particles are also
discussed, which can be used to categorize and characterize MRJ
packings according to the particle shapes and symmetries.


The rest of the paper is organized as follows: In Sec. II, we
provide preliminaries and basic definitions pertinent to packing problems.
In Sec. III, we formulate general organizing principles for dense
packings of nonspherical hard particles in the form of four
different propositions and test them against the most
comprehensive set of the densest known packing constructions of
both convex and concave nonspherical particles known to date. In
Sec. IV, we present and discuss additional applications of our
general organizing principles. Specifically, we construct the
densest known packings of centrally symmetric spherocylinders,
``lens-shaped'' particles, as well as non-centrally symmetric
square and rhombic pyramids. We also discuss the expected
high-density equilibrium crystal phases of a variety of
nonspherical particles as predicted by our propositions. In Sec.
V, we make concluding remarks and briefly discuss the unique
characteristics of MRJ packings.

\section{Preliminaries and Definitions}



In this section, we provide some basic definitions concerning
packings, which closely follow those given in Ref.~\cite{ToStRMP}.
A {\it packing} $P$ is a collection of nonoverlapping solid
objects or particles in $d$-dimensional Euclidean space
$\mathbb{R}^d$. Packings can be defined in other spaces (e.g.,
hyperbolic spaces and compact spaces, such as the surface of a
$d$-dimensional sphere), but our primary focus in this review is
$\mathbb{R}^d$. A {\it saturated} packing is one in which there is
no space available to add another particle of the same kind to the
packing. A \textit{uniform} packing has a symmetry operation
(e.g., a point inversion-symmetric transformation) that takes any
particle into another.

\begin{figure}[htbp]
\begin{center}
$\begin{array}{c@{\hspace{1.0cm}}c}
\includegraphics[width=5.5cm,keepaspectratio]{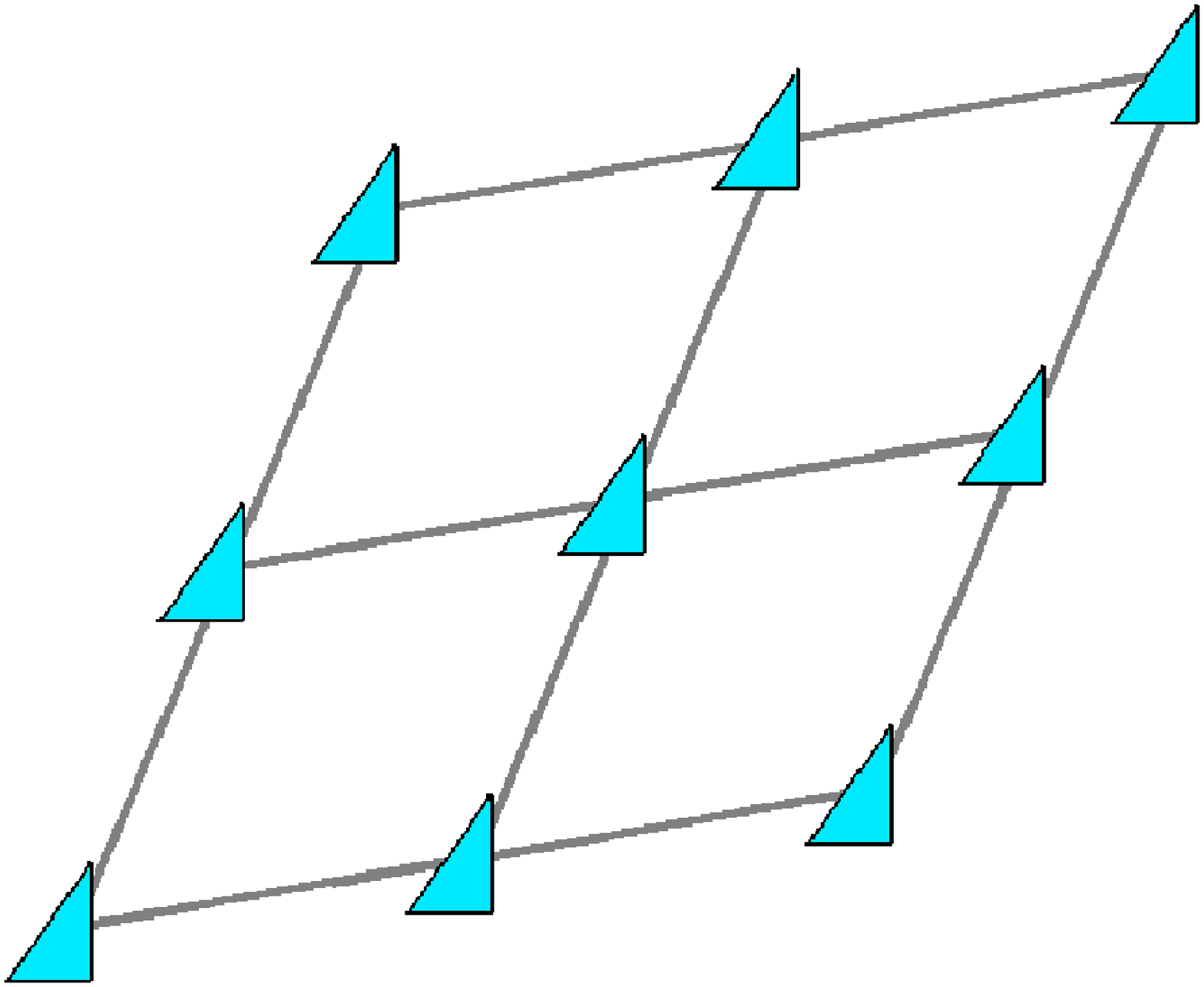} &
\includegraphics[width=5.5cm,keepaspectratio]{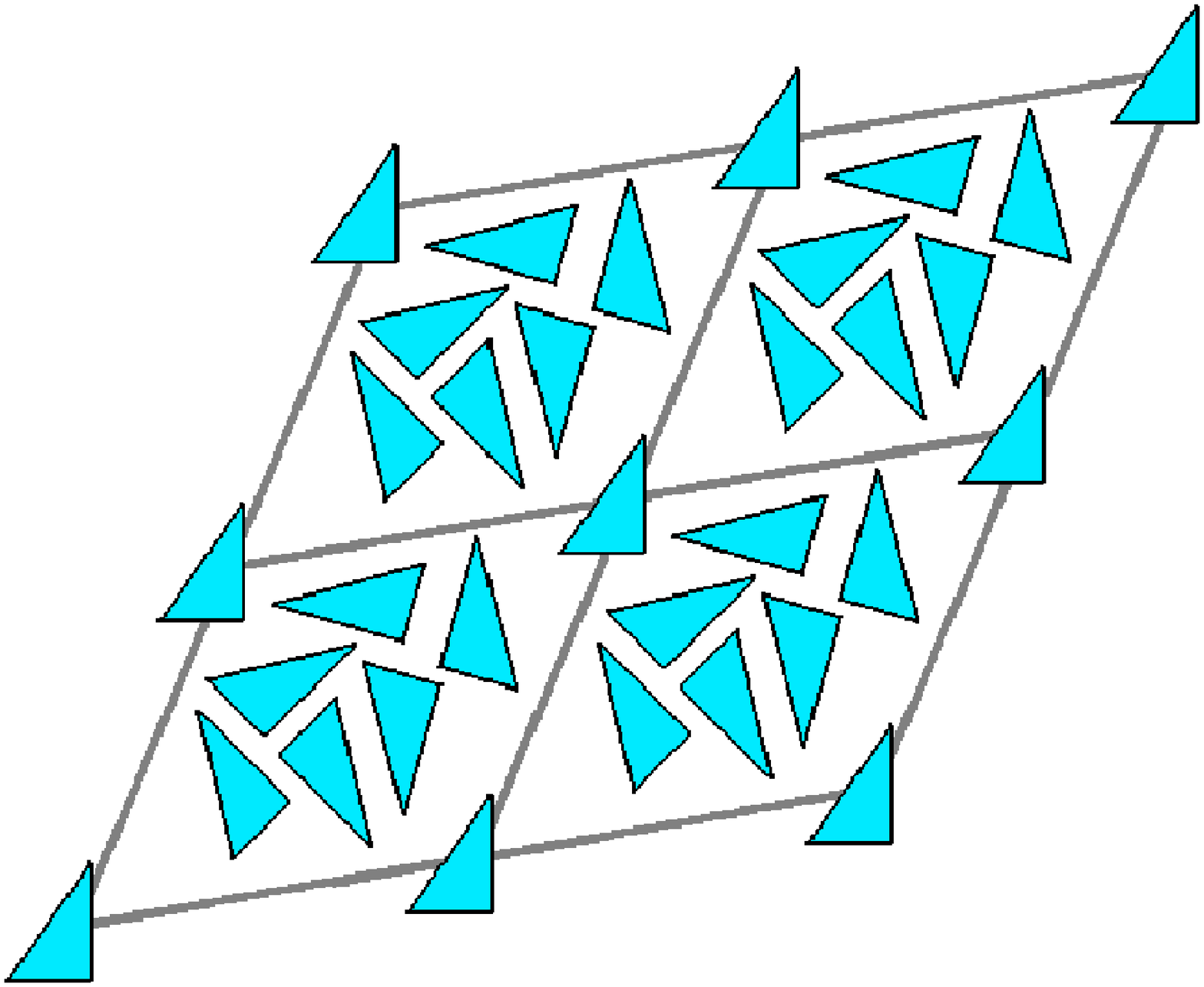} \\
\mbox{(a)} & \mbox{(b)} \\
\end{array}$
\end{center}
\caption{(Color online) (a) A portion of a lattice packing of congruent
nonspherical particles. Each fundamental cell (depicted as a
rhombus here) has exactly one particle centroid. Each particle in
the packing must have the same orientation. (b) A portion of a
periodic non-lattice packing of congruent nonspherical particles.
The fundamental cell contains  multiple nonspherical particles
with arbitrary positions and orientations within the cell subject
to the nonoverlap constraint.} \label{lat-period}
\end{figure}

A {\it lattice} $\Lambda$ in $\mathbb{R}^d$ is a subgroup
consisting of the integer linear combinations of vectors that
constitute a basis for $\mathbb{R}^d$. In the physical sciences
and engineering, this is referred to as a {\it Bravais} lattice.
Unless otherwise stated, the term ``lattice'' will refer here to a
Bravais lattice only. A {\it lattice packing} $P_L$ is one in
which  the centroids of the nonoverlapping identical particles are
located at the points of $\Lambda$, and all particles have a
common orientation. The set of lattice packings is a subset of all
possible packings in $\mathbb{R}^d$. In a lattice packing, the
space $\mathbb{R}^d$ can be geometrically divided into identical
regions $F$ called {\it fundamental cells}, each of which contains
the centroid of just one particle. Thus, the density of a lattice
packing is given by
\begin{equation}
\phi= \frac{v_1}{\mbox{Vol}(F)},
\end{equation}
where $v_1$ is the volume of a single $d$-dimensional particle and
$\mbox{Vol}(F)$ is the $d$-dimensional volume of the fundamental
cell. For example, the volume $v_1(R)$ of  a $d$-dimensional
spherical particle of radius $R$ is given explicitly by
\begin{equation}
v_1(R)=\frac{\pi^{d/2} R^d}{\Gamma(1+d/2)},
\label{vol-sph}
\end{equation}
where $\Gamma(x)$ is the Euler gamma function. Figure
\ref{lat-period}(a) depicts lattice packings of congruent spheres
and congruent nonspherical particles.

A more general notion than a lattice packing is a periodic
packing. A {\it periodic} packing of congruent particles is
obtained by placing a fixed configuration of $N$ particles (where
$N\ge 1$) with {\it arbitrary nonoverlapping orientations} in one
fundamental cell of a lattice $\Lambda$, which is then
periodically replicated without overlaps. Thus, the packing is
still periodic under translations by $\Lambda$, but the $N$
particles can occur anywhere in the chosen fundamental cell
subject to the overall nonoverlap condition. The packing density
of a  periodic packing is given by
\begin{equation}
\phi=\frac{N v_1}{\mbox{Vol}(F)}=\rho v_1,
\end{equation}
where $\rho=N/\mbox{Vol}(F)$ is the number density, i.e., the
number of particles per unit volume. Figure \ref{lat-period}(b)
depicts a periodic non-lattice packing of congruent spheres and
congruent nonspherical particles. Note that the particle
orientations within a fundamental cell in the latter case are
generally not identical to one another.

\begin{figure}[htbp]
\begin{center}
$\begin{array}{c@{\hspace{0.6cm}}c@{\hspace{0.6cm}}c}
\includegraphics[width=3.5cm,keepaspectratio]{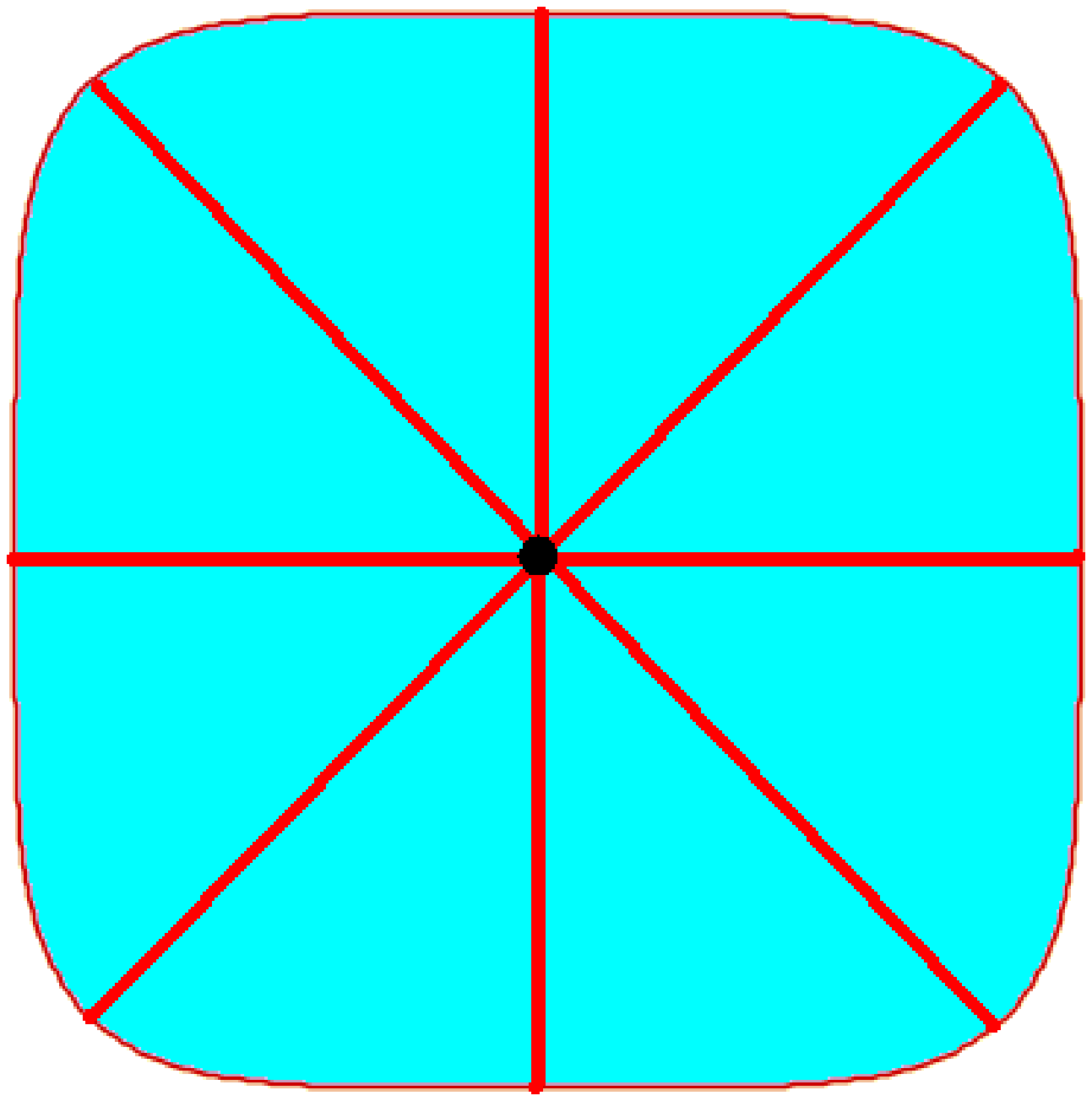} &
\includegraphics[width=3.5cm,keepaspectratio]{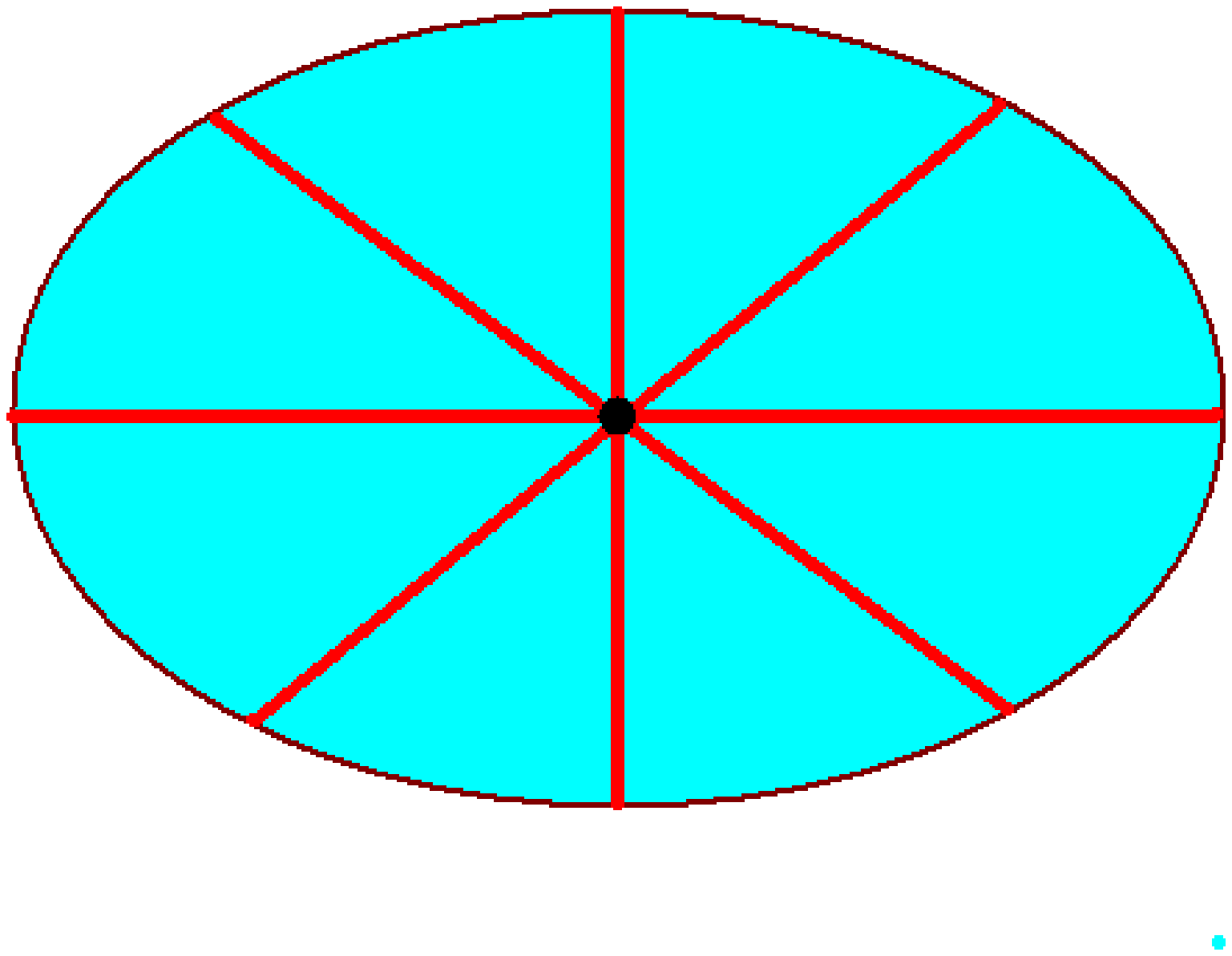} &
\includegraphics[width=3.5cm,keepaspectratio]{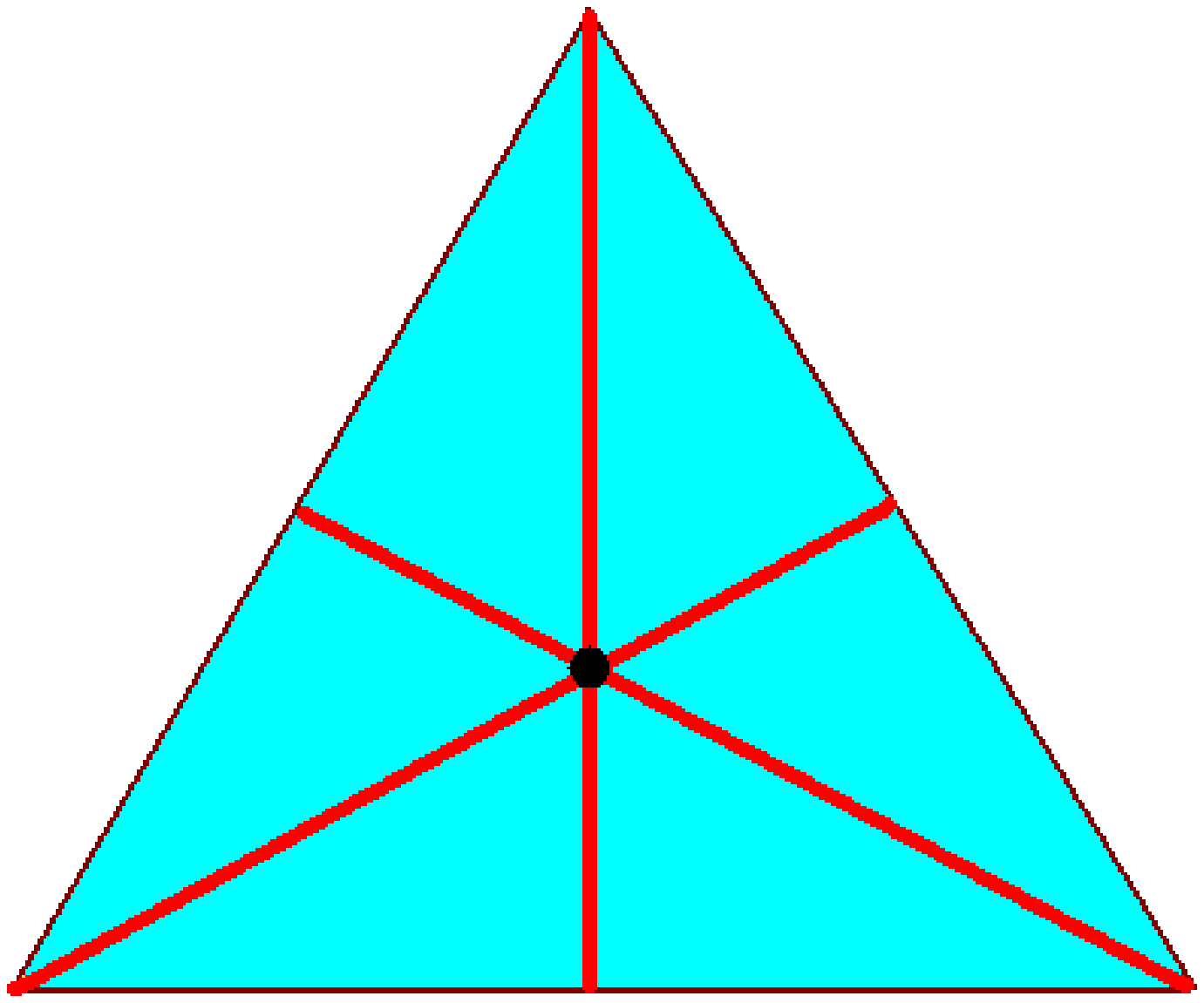} \\
\mbox{(a)} & \mbox{(b)} & \mbox{(c)} \\
\end{array}$
\end{center}
\caption{(Color online) (a) A ``superdisk" is centrally symmetric and possesses
two equivalent principal axes. (b) An ellipse  is centrally
symmetric but does not possess two equivalent principal axes. (c)
A triangle is not centrally symmetric. } \label{central}
\end{figure}

We will see subsequently that certain characteristics of particle
shape, e.g., whether it possesses central symmetry and equivalent
principal axes, play a fundamental role in determining its dense
packing configurations. A $d$-dimensional particle is {\it
centrally symmetric} if it has a center $C$ that bisects every
chord through $C$ connecting any two boundary points of the
particle, i.e., the center is a point of inversion symmetry.
Examples of centrally symmetric particles in  $\mathbb{R}^d$
include spheres, ellipsoids and superballs (see
Ref.~\cite{ToStRMP} for definitions of these shapes). A triangle
and tetrahedron are examples of non-centrally symmetric two- and
three-dimensional particles, respectively. Figure \ref{central}
depicts examples of centrally and non-centrally symmetric
two-dimensional particles. A $d$-dimensional centrally symmetric
particle for $d \ge 2$ is said to possess $d$ equivalent principal
(orthogonal) axes (directions) associated with the moment of
inertia tensor if those directions are two-fold rotational
symmetry axes such that the $d$ chords along those directions and
connecting the respective pair of particle-boundary points are
equal. (For $d=2$, the two-fold (out-of-plane) rotation along an
orthogonal axis brings the shape to itself, implying the rotation
axis is a ``mirror image'' axis.) Whereas a $d$-dimensional
superball has $d$ equivalent directions, a $d$-dimensional
ellipsoid generally does not (see Fig. \ref{central}).

\section{General Organizing Principles}

We have formulated several organizing principles in the form of
three conjectures for convex polyhedra as well as other
nonspherical convex shapes \cite{ToJi09, ToJi09b, ToJi10,
TrunTetrah}; see Fig. 5 for
examples of dense packings of certain convex nonspherical hard particles.
In this section, we generalize them in order to guide
one to ascertain the densest packings of other convex nonspherical
particles as well as {\it concave} shapes based on the
characteristics of the particle shape (e.g., symmetry, principal
axes and local principal curvature). The generalized organizing
principles are explicitly stated as four distinct propositions. We
apply and test all of these organizing principles to the most
comprehensive set of both convex and concave particle shapes examined to date,
including Catalan solids, prisms, antiprisms, cylinders, dimers of
spheres and various concave polyhedra. We demonstrate that all of
the densest known packings associated with this wide spectrum of
nonspherical particles are consistent with our propositions. In
Sec. IV, we will apply our propositions to construct analytically
the densest known packings of other nonspherical particles,
including spherocylinders and ``lens-shaped'' particles that are
centrally symmetric, as well as square and rhombic pyramids that
lack central symmetry. We also apply the organizing principles to
infer the high-density equilibrium crystalline phases of hard
convex and concave particles in Sec. IV.

\subsection{Organizing Principles for Convex Particles}

\begin{figure}[htbp]
\begin{center}
$\begin{array}{c@{\hspace{0.6cm}}c@{\hspace{0.6cm}}c}
\includegraphics[width=4.5cm,keepaspectratio]{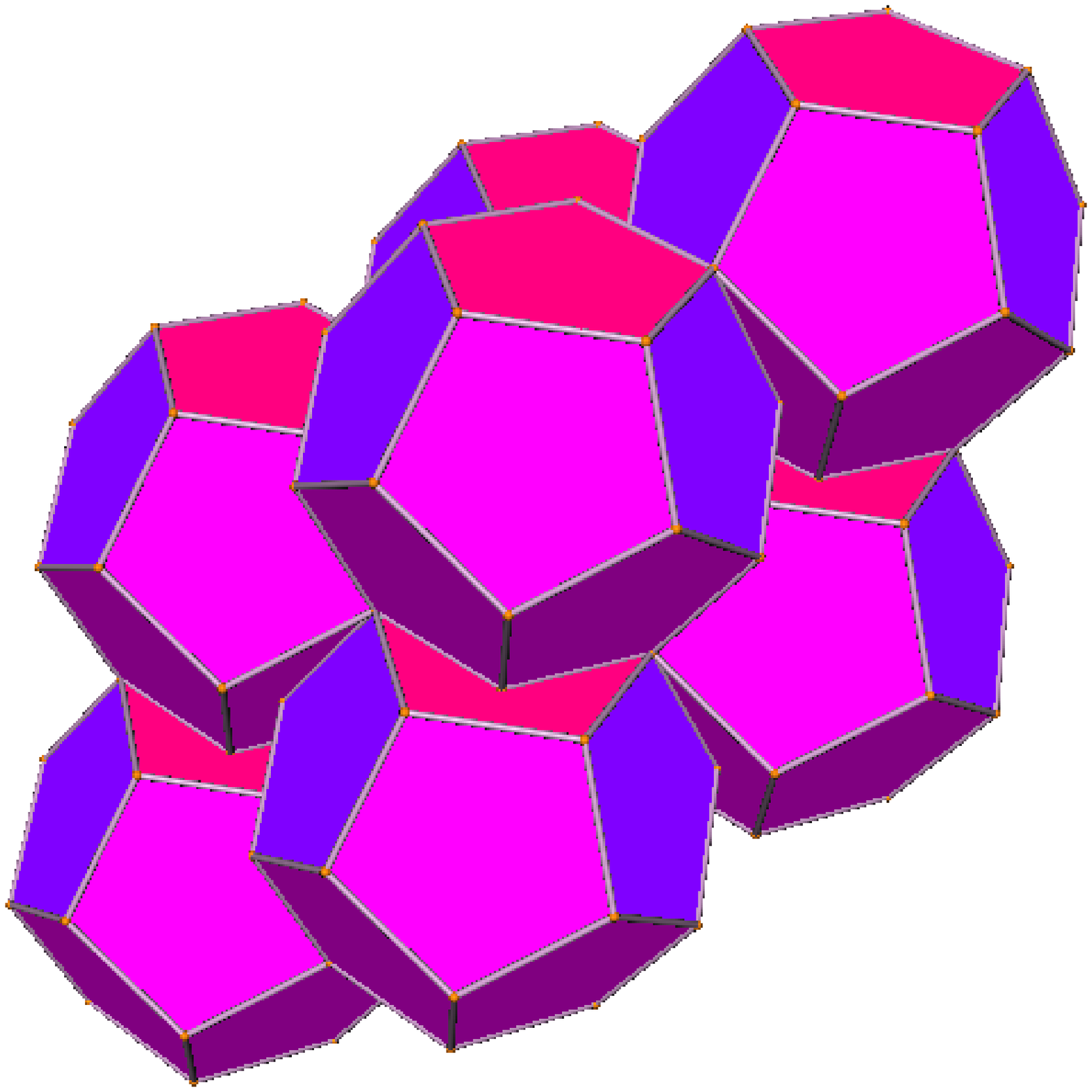} &
\includegraphics[width=4.5cm,keepaspectratio]{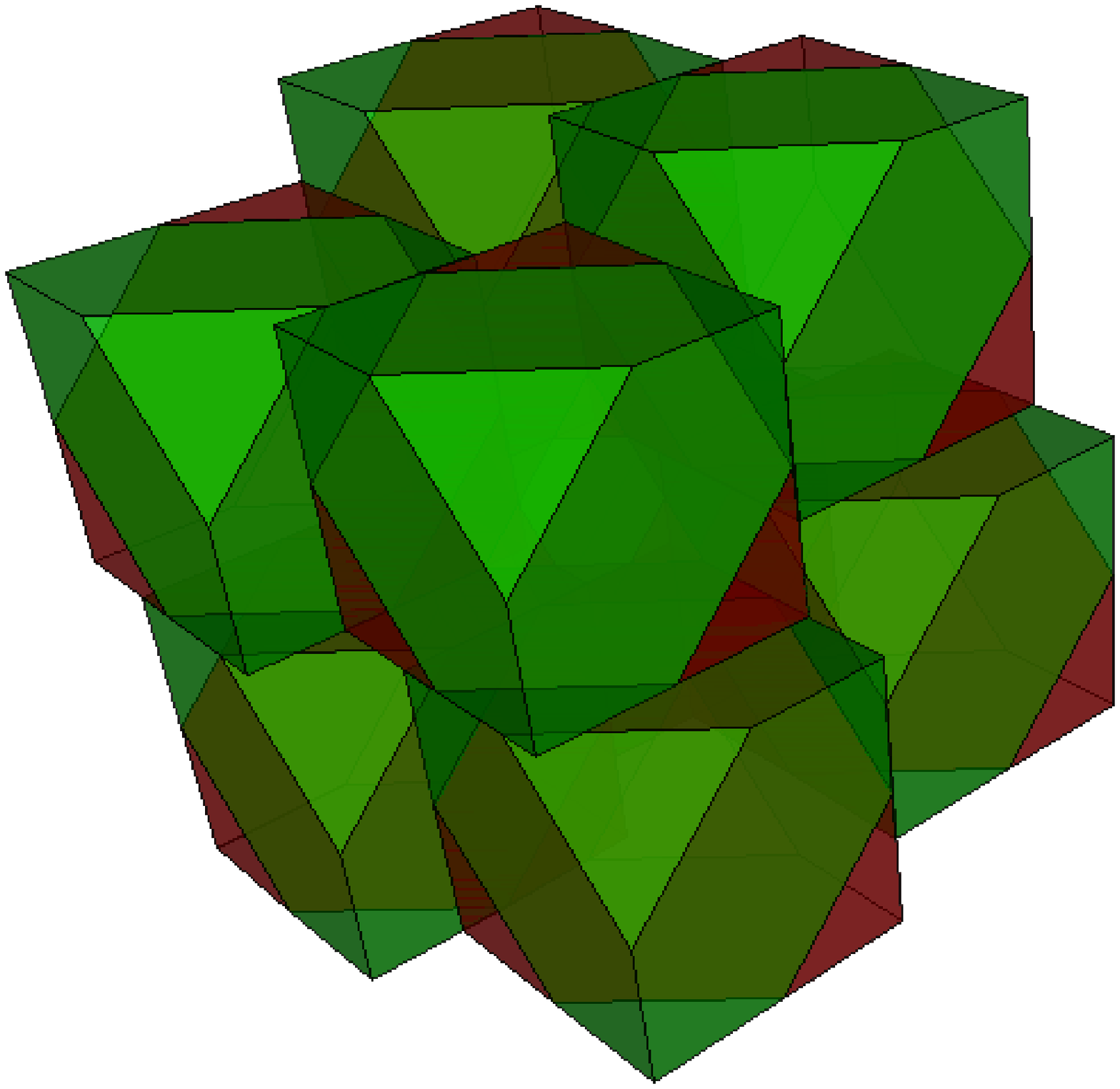} &
\includegraphics[width=4.5cm,keepaspectratio]{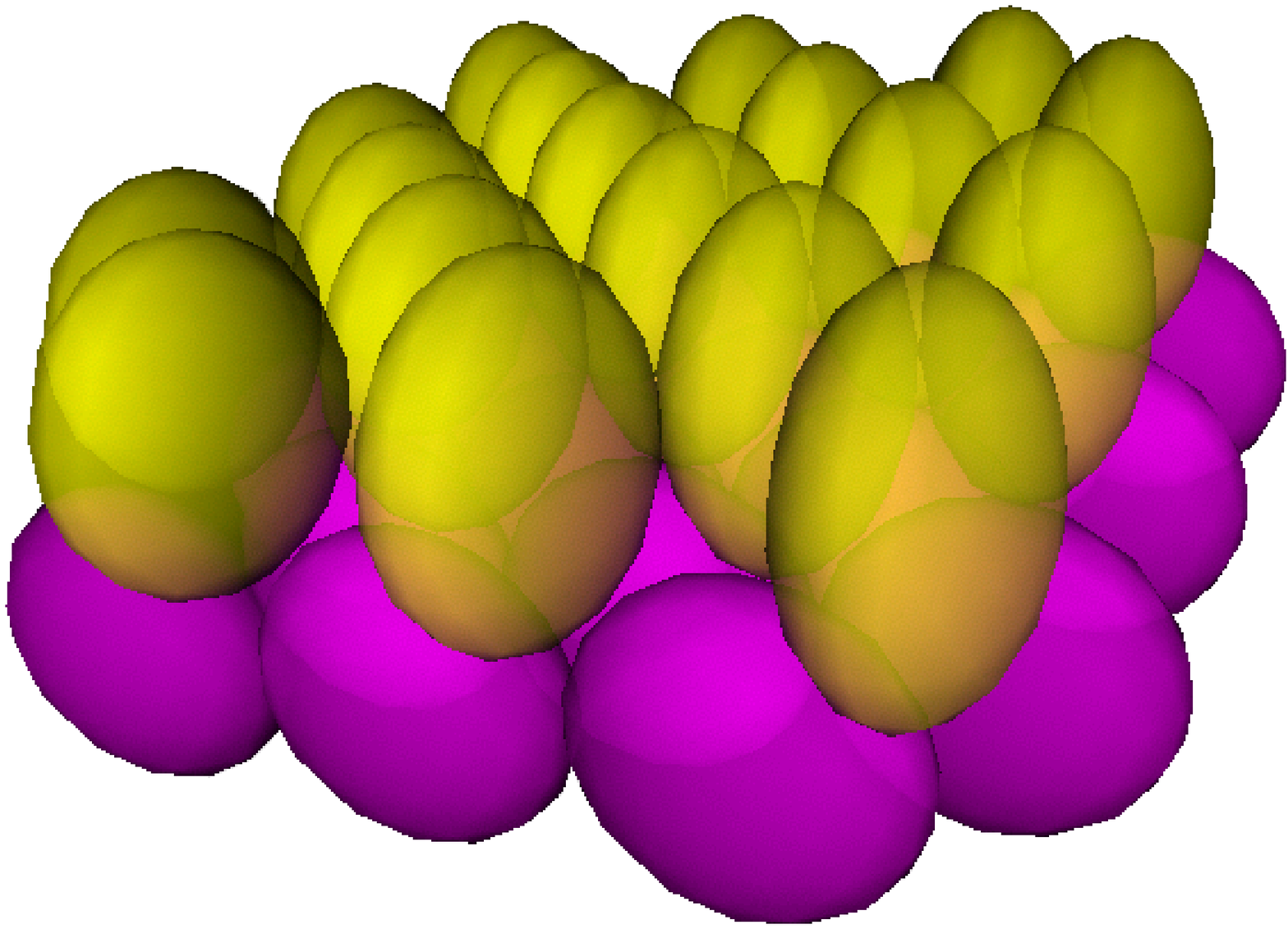} \\
{\mbox (a)} & {\mbox (b)} & {\mbox (c)} \\
\end{array}$
\end{center}
\caption{(Color online) Dense packings of nonspherical particles. (a) A portion
of the densest lattice packing of dodecahedra, which possess
central symmetry. (b) A portion of the densest known packing of
truncated tetrahedra without central symmetry. In the packing, the
truncated tetrahedra form centrally symmetric dimers and the
dimers pack on a lattice. (c) A portion of the densest known
packing of oblate ellipsoids, which contains two ellipsoids perpendicular
to one another in the fundamental cell.} \label{convex_packing}
\end{figure}

For the centrally symmetric Platonic and Archimedean solids, our
simulations and theoretical analysis lead us to the following
conjecture \cite{ToJi09, ToJi09b}:

\bigskip
\noindent{\bf Conjecture 1}: {\sl The densest
packings of the centrally symmetric Platonic and Archimedean
solids are given by their corresponding optimal Bravais lattice
packings.}
\bigskip


\noindent {\it Remarks}: The centrally symmetric Platonic and
Archimedean solids are polyhedra with three equivalent principal
axes. For packings of polyhedra, face-to-face contacts
are favored over other types of contacts (e.g., face-to-edge,
face-to-vertex, edge-to-edge etc.), since the former allow the
particle centroids to get closer to one another and thus, to
achieve higher packing densities. For centrally symmetric
polyhedra, aligning the particles (i.e., all in the same
orientation) enable a larger number of face-to-face contacts. For
example, a particle with $F$ faces, possesses $F/2$ families of
axes that go through  the centroid of the particle and intersect
the centrally-symmetric face pairs such that the particles (in the
same orientation) with their centroids arranged on these axes can
form face-to-face contacts. The requirement that the particles
have the same orientation is globally consistent with a (Bravais)
lattice packing. Indeed, in the optimal lattice packings of the
centrally symmetric Platonic and Archimedean solids, each particle
has the maximum number of face-to-face contacts that could
possibly be obtained without violating the nonoverlapping
conditions. It is highly unlikely that such particles possessing
three equivalent principal axes and aligned in the same direction
could form a more complicated non-lattice periodic packings with
densities that are larger than the optimal lattice packings.

\begin{table}[htbp]
\caption{Characteristics of dense packings of Platonic and
Archimedean solids, including whether the particle possesses
central symmetry, whether the packing is a Bravais lattice packing, the
number of basis particles in the fundamental cell if the packing is
a non-Bravais-lattice periodic packing, the numerically
obtained packing density $\phi^*$ \cite{ToJi09, ToJi09b,
Dijkstra}, the analytically obtained density $\phi_{max}$
\cite{Be00}, and the upper bound on the density $\phi^U_{max}$
\cite{ToJi09, ToJi09b}.} \label{tab_poly}
\begin{tabular}{c@{\hspace{0.1cm}}c@{\hspace{0.3cm}}c@{\hspace{0.3cm}}c@{\hspace{0.3cm}}c@{\hspace{0.3cm}}c}
\hline\hline
Polyhedron & Cen. Sym. & Structure & $\phi^*$ & $\phi_{max}$ & $\phi^U_{max}$ \\
\hline
Tetrahedron & No & Periodic, 4-particle basis & 0.8560 & 0.856347 & 1 \\
Icosahedron & Yes & Bravais Lattice & 0.8361 & 0.836357 & 0.893417 \\
Dodecahedron & Yes & Bravais Lattice & 0.9042 & 0.904508 & 0.981162 \\
Octahedron & Yes & Bravais Lattice & 0.9471 & 0.947368 & 1 \\
Cube & Yes & Bravais Lattice & 0.9999 & 1 & 1 \\
Truncated Tetrahedron & No & Periodic, 2-particle basis & 0.9885 & 0.995192 & 1 \\
Truncated Icosahedron & Yes & Bravais Lattice & 0.7849 & 0.784987 & 0.838563 \\
Snub Cube & No & Bravais Lattice & 0.7876 & 0.787699 & 0.934921 \\
Snub Dodecahedron & No & Bravais Lattice & 0.7874 & $0.788640$ & $0.855474$ \\
Rhombicicosidodecahedron & Yes & Bravais Lattice & 0.8047 & $0.804708$ & $0.835964$ \\
Truncated Icosidodecahedron & Yes & Bravais Lattice & 0.8269 & $0.827213$ & $0.897316$ \\
Truncated Cuboctahedron & Yes & Bravais Lattice & 0.8489 & $0.849373$ & $0.875805$ \\
Icosidodecahedron & Yes & Bravais Lattice & 0.8644 & $0.864720$ & $0.938002$ \\
Rhombicuboctahedron & Yes & Bravais Lattice & 0.8758 & 0.875805 & 1 \\
Truncated Dodecahedron & Yes & Bravais Lattice & 0.8976 & $0.897787$ & $0.973871$ \\
Cuboctahedron & Yes & Bravais Lattice & 0.9182 & $0.918367$ & 1 \\
Truncated Cube & Yes & Bravais Lattice & 0.9722 & $0.973747$ & 1 \\
Truncated Octahedron & Yes & Bravais Lattice & 0.9999 & 1 & 1 \\
\hline\hline
\end{tabular}

\end{table}

\begin{table}[htbp]
\caption{Characteristics of dense packings of Catalan solids and
certain prisms, including whether the particle possesses central
symmetry, whether the packing is a Bravais lattice packing, the number of
basis particles in the fundamental cell if the packing is
a non-Bravais-lattice periodic packing \cite{Dijkstra}, the
numerically obtained packing density $\phi^*$ \cite{Dijkstra}, and
the upper bound on the density $\phi^U_{max}$ \cite{Dijkstra}.}
\label{tab_poly_cata}
\begin{tabular}{c@{\hspace{0.1cm}}c@{\hspace{0.4cm}}c@{\hspace{0.4cm}}c@{\hspace{0.4cm}}c}
\hline\hline
Polyhedron & Cen. Sym. & Structure & $\phi^*$ & $\phi^U_{max}$ \\
\hline
Pentagonal Hexecontrahedron & No & Bravais Lattice & 0.741075 & 0.782834 \\
Pentagonal Icositetrahedron & No & Periodic, 2-particle basis & 0.743639  & 0.848563 \\
Pentakis Dodecahedron & Yes & Bravais Lattice & 0.757552  & 0.787991 \\
Disdyakis Triacontrahedron & Yes & Bravais Lattice & 0.765496  & 0.773134 \\
Deltoidal Hexecontrahedron & Yes & Bravais Lattice & 0.77155  & 0.782871 \\
Disdyakis Dodecahedon & Yes & Bravais Lattice & 0.793288  & 0.813656 \\
Deltoidal Icositetrahedron & Yes & Bravais Lattice & 0.796934  & 0.851348 \\
Triakis Tetrahedron & No & Periodic, 4-particle basis & 0.798868  & 1 \\
Rhombic Triacontrahedron & Yes & Bravais Lattice & 0.801741  & 0.834626 \\
Triakis Icosahedron & Yes & Bravais Lattice & 0.804796  & 0.818047 \\
Tetrakis Hexahedron & Yes & Bravais Lattice & 0.814019  & 0.878410 \\
Small Triakis Octahedron & Yes & Bravais Lattice & 0.876016  & 0.937283 \\
Rhombic Dodecahedron & Yes & Bravais Lattice & 1  & 1 \\
Heptaprism & No & Periodic, 2-particle basis & 0.896  & 1 \\
Pentaprism & No & Periodic, 2-particle basis & 0.921  & 1 \\
\hline\hline
\end{tabular}
\end{table}

Conjecture 1 is the analog of Kepler's sphere conjecture for the
centrally symmetric Platonic and Archimedean solids. In this
sense, such  solids behave similarly to spheres in that their
densest packings are lattice arrangements and (except for the
cube) are geometrically frustrated like spheres. This conjecture
was strongly supported by our original simulations of the Platonic
solids \cite{ToJi09, ToJi09b}, which used multiple-particle
configurations in the fundamental cell and only produce the
optimal lattice packings. More recently, dense packings of other
polyhedra have been numerically investigated, including the
Archimedean solids \cite{Dijkstra, TrunTetrah}, Catalan solids
\cite{Dijkstra}, Johnson solids \cite{Dijkstra}, prisms and
antiprisms \cite{Dijkstra}, and a family of truncated tetrahedra
\cite{GlozterTrunTetrah}. Among this large set of polyhedra, it
has been found that for centrally symmetric shapes, numerical
simulations with multiple particles in the fundamental cell always
converge to dense lattice packings, which possess higher densities
than all other periodic packing configurations. These findings
support Conjecture 1. Certain characteristics of the densest known
packings of a subset of the aforementioned polyhedra are provided
in Table~I and Table~II, including the type of the packing
structure (lattice or periodic) and the packing density (both
numerical value and analytical value if known).

In addition to the strong numerical evidence for Conjecture 1, we
have derived a simple {\it analytical} upper bound on the maximal
density $\phi_{max}$ of a packing of congruent nonspherical
particles of volume $v_p$ in any Euclidean space dimension $d$ in
terms of  the maximal density of a $d$-dimensional packing of
congruent spheres of volume $v_s$, which is the volume of the
largest sphere than can be inscribed in the nonspherical particle.
Here, for simplicity, we state that upper bound for
three-dimensions:

\bigskip
\noindent{{\bf Lemma}:} {\sl The maximal density of a packing of
congruent nonspherical particles is bounded from above according
to the following bound \cite{ToJi09, ToJi09b}
\begin{equation}
\phi_{max}\le \mbox{min}\left[\frac{v_p}{v_s} \frac{\pi}{\sqrt{18}},~1\right],
\label{lemma}
\end{equation}
where $\mbox{min}[x,y]$ denotes the minimum of $x$ and $y$.}
\bigskip

The upper bound values for a subset of the aforementioned
polyhedra are give in Table~I and Table~II. As noted in
Ref.~\cite{ToJi09b}, the upper bound (\ref{lemma}) is generally
not sharp, i.e., it generally can not be realized, except for
certain space-filling shapes. Therefore, a small difference
between the actual packing density and the corresponding upper
bound value strongly suggests the optimality of the packing. It
can be seen from Tables~I and II that the difference is indeed
small for centrally symmetric polyhedra with equivalent principal
axes. Together with the numerical evidence, this leads us to the
following more general proposition:

\bigskip
\noindent{\bf Proposition 1}: {\sl Dense packings of centrally
symmetric convex, congruent polyhedra with three equivalent axes
are given by their corresponding densest lattice packings,
providing a tight density lower bound that may be optimal.}
\bigskip

\noindent {\it Remarks}: This proposition is the three-dimensional
analog of the Fejes-T{\'o}th theorem for two-dimensional centrally
symmetric shapes \cite{Toth}. Note that whenever Proposition
1 leads to an optimal packing, one can immediately obtain an upper
bound on $\phi_{max}$ that improves upon inequality (\ref{lemma}).
Namely, for a general nonspherical particle of volume $v_p$,
instead of inscribing into the particle the largest possible
sphere, one can inscribe the largest possible polyhedron with known
maximal packing density, i.e.,
\begin{equation}
\phi_{max}\le \mbox{min}\left[\frac{v_p}{v_0}
\phi^0_{max},~1\right], \label{lemma2}
\end{equation}
where $v_0$ denotes the volume of the largest centrally symmetric
convex particle with three equivalent axes that can be inscribed
in the nonspherical particle of interest and $\phi^0_{max}$ is its
associated maximal density. In the case that $\phi^0_{max}$ can
be proven to be optimal, (\ref{lemma2}) is a rigorous upper bound.
Otherwise, it provides an estimate of $\phi_{max}$.

As an application of inequality (5), consider inscribing a
truncated dodecahedron into a dodecahedron. Assuming that
$\phi^0_{max} = 0.897787$ for the optimal truncated dodecahedron
packing leads to an estimate of $\phi^* = 0.904508$ for the
densest dodecahedron packing, which is identical to the actual
value $\phi_{max} = 0.904508$ for the densest {\it lattice}
packing of dodecahedra. In general, the inscribed polyhedron
should be chosen carefully in order to get a nontrivial upper
bound. In the case that the optimality of the packing of the
inscribed polyhedron can be proven, this could lead to tight upper
bound on $\phi_{max}$ for nonspherical particles improved over the
simple sphere bound (\ref{lemma}).

Note that the space-filling polyhedra in the aforementioned three
families of shapes, i.e., the cube, the truncated octahedron (one
of the Archimedean solids) and the rhombic dodecahedron (dual of
the Catalan cuboctahedron) are respectively the Voronoi cells
associated with the simple cubic, body-centered cubic and
face-centered cubic lattices \cite{SalBook}. The densest packings
of cubes can be achieved by an uncountably infinite number of
non-lattice packings corresponding to sliding layers or chains of
cubes in their simple-cubic-lattice packing \cite{ToJi09b}. By
contrast, we note here that the densest Bravais-lattice packings
of the truncated octahedron and the rhombic dodecahedron are the
unique densest packings of the two shapes. In contrast to the
densest lattice packing of cubes, in the densest lattice packings
of the truncated octahedron and the rhombic dodecahedron, no
chains or layers of particles can slide relative to one another.
Thus, there are no collective motions within the densest packings
\cite{tiling} of the truncated octahedron and the rhombic
dodecahedron that can lead to other non-lattice packings of the
two shapes that can completely fill space.

We also note that both simulations \cite{JiaoSuperball,
NiSuperball} and analytical constructions \cite{JiaoSuperball}
suggest that the optimal packings of superballs, a rich family of
smoothly-shaped centrally symmetric particles with both octahedral
and cubic symmetry including a three-dimensional cross
limit, are their densest Bravais lattice packings. This suggests
that Conjecture 1 can be generalized not only to polyhedral
particles but also to certain smoothly-shaped particles. A major
distinction between a smoothly-shaped particle and a polyhedron is
that the surface of the former possesses non-trivial local
principal curvatures, which are either zero or infinity for a polyhedron. We note
that a complete understanding of the role of curvature in
determining dense packing configurations is currently lacking,
although it has been shown that curvature is crucial for
stabilized disordered packings of nonspherical particles
\cite{DoSiSaVa04, JiStTo10}.


For convex polyhedral particles that lack central symmetry, we
have made the following conjecture concerning their densest
packing configurations \cite{ToJi09, ToJi09b}:

\bigskip
\noindent{\bf Conjecture 2}: {\sl The densest packing of any
convex, congruent polyhedron without central symmetry generally is
not a Bravais lattice packing, i.e., set of such polyhedra whose
optimal packing is not a Bravais lattice is overwhelmingly larger
than the set whose optimal packing is a Bravais lattice. }
\bigskip


\noindent {\it Remarks}: In other words, Conjecture 2 states that
the set of non-centrally symmetric polyhedra whose optimal packing
is not a lattice is overwhelmingly larger than the set whose
optimal packing is a lattice. As shown in Table~I and Table~II,
the fact that the regular tetrahedron and truncated tetrahedron
lack central symmetry and that dense packings of such objects
favor face-to-face contacts immediately eliminates the possibility
that lattice packings (in which particles must have the same
orientations and thus, have face-to-vertex contacts) are optimal.
Similarly, it is very plausible that dense packings of most
convex, congruent polyhedra without central symmetry are
facilitated by face-to-face contacts by exploring their rotational
degrees of freedom and hence the optimal packings are generally
not lattice packings. This reasoning immediately leads us to the
following more general proposition concerning dense packings of
polyhedra without central symmetry:

\bigskip
\noindent{\bf Proposition 2}: {\sl Dense packings of convex,
congruent polyhedra without central symmetry are composed of
centrally symmetric compound units of the polyhedra with the
inversion-symmetric points lying on the densest lattice associated
with the compound units, providing a tight density lower bound
that may be optimal.}
\bigskip

\begin{figure}[htbp]
\begin{center}
$\begin{array}{c@{\hspace{1.0cm}}c}\\
\includegraphics[height=4.5cm,keepaspectratio]{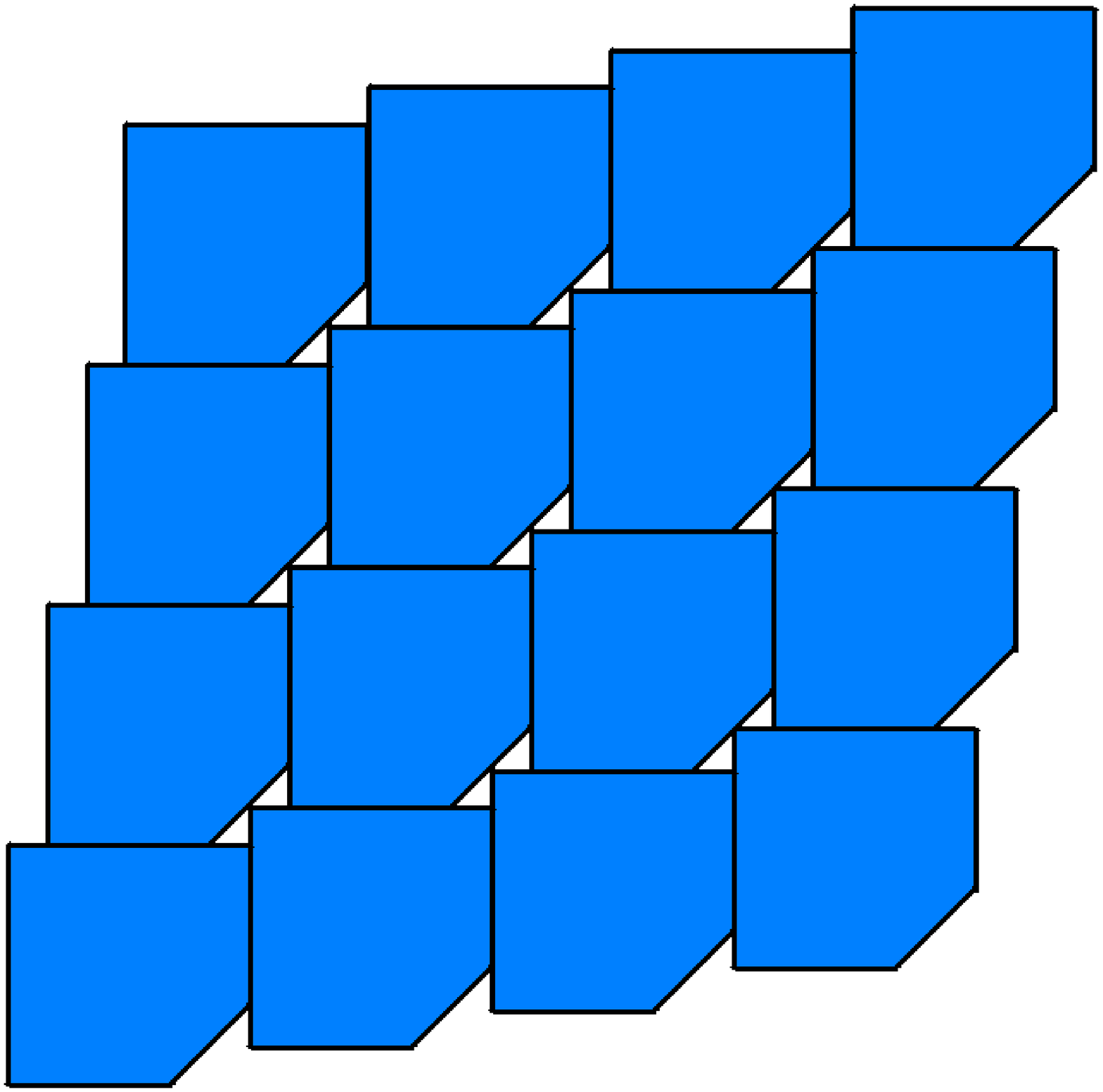} &
\includegraphics[height=4.35cm,keepaspectratio]{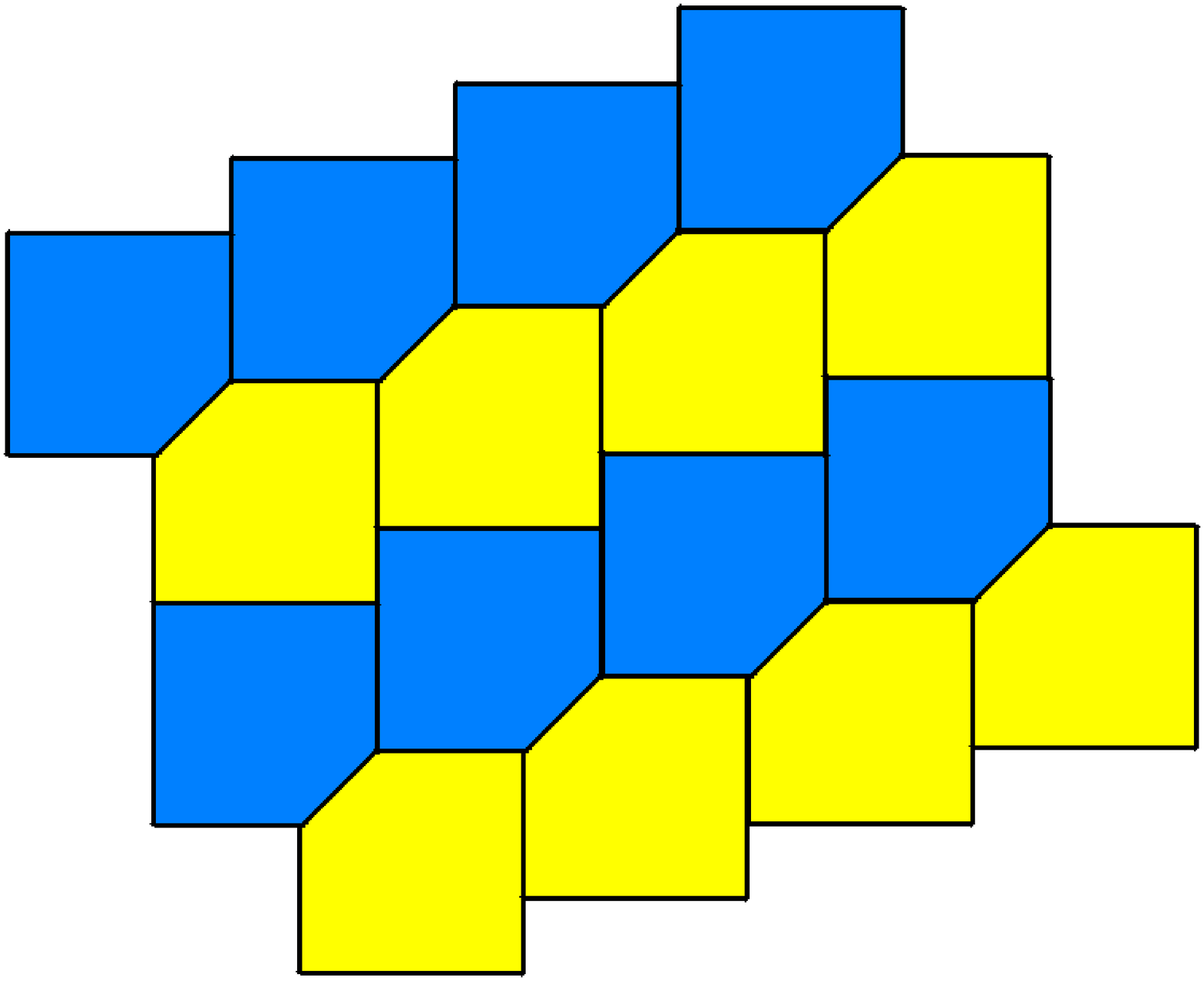} \\
\mbox{(a)} & \mbox{(b)}
\end{array}$
\caption{(Color online) Portions of two packing configurations of
pentagons obtained by cutting off a corner (isosceles triangle
with a right angle) of a square (adapted from
Ref.~\cite{ToJi09b}). (a) The optimal lattice packing and (b) a
two-particle basis periodic packing that tiles the plane. Let side
length of the square be 1 and the lengths of the equal sides of
the isosceles triangle be $\beta \in (0,1)$. The lattice vectors
of the optimal lattice packing are ${\bf e}^L_1 = {\bf i} -
\frac{\beta}{2} {\bf j}$, ${\bf e}^L_2 = (1-\frac{\beta}{2}){\bf
i} + (1-\beta) {\bf j}$ and the lattice vectors of the periodic
packing are ${\bf e}^P_1 = {\bf i} + \beta {\bf j}$, ${\bf e}^P_2
= (1-\beta){\bf i} + (2-\beta) {\bf j}$ with one particle at
origin and the other at ${\bf b}_1 = (1-\beta){\bf
i}+(1-\beta){\bf j}$, where ${\bf i}, {\bf j}$ are the unit
vectors along the two orthogonal coordinate directions, coinciding
with two orthogonal sides of the square. The density (covering
fraction) of the optimal lattice packing is $\phi_{max}^L =
(1-\frac{\beta^2}{2})/(1-\frac{\beta^2}{4})$ and the density of
the periodic packing is $\phi_{max}^P = 1$. It can be seen that
for all $0<\beta<1$, $\phi_{max}^L$ is always smaller than
$\phi_{max}^P$. } \label{Cut_Square}
\end{center}
\end{figure}

\noindent {\it Remarks}: Application of this proposition can be
very well illustrated by the ``truncated square'' example
originally given in Ref.~\cite{ToJi09b}. We note that
although this specific example involves polygons in two
dimensions, the general idea that lacking central symmetry enables
the full exploration of the rotational degrees of freedom
associated with a nonspherical particle in obtaining the densest
packing also applies in three dimensions. Specifically, consider
a square with one missing corner, i.e., an isosceles triangle with
a right angle (see Fig.~\ref{Cut_Square}). At first glance, one
might surmise that if the missing piece is sufficiently small, a
lattice packing which is close to the original square lattice
packing should still be optimal (see Fig.~\ref{Cut_Square}a),
since lattice packings are optimal for squares. However, no matter
how small the missing piece may be, a periodic packing in which
the fundamental cell contains two truncated squares can be
constructed that tile the plane (see Fig.~\ref{Cut_Square}b). This
is done by taking advantage of the asymmetry of the particle to
maximize possible face-to-face contacts. Thus, we see from this
counterintuitive example that if the particle does not possess
central symmetry, it is possible to exploit its rotational degrees
of freedom to yield a periodic packing with a complex basis that
are generally denser than the optimal lattice packing. We also
note that Proposition 2 can be considered as the three-dimensional
analog of Kuperberg-Kuperberg double-lattice packing constructions
for two-dimensional convex shapes lacking central symmetry
\cite{kuper2}.

On the other hand, there are special cases where the lattice will
be optimal for particles lacking central symmetry. One such
example in three dimensions is the rhombic dodecahedron that has
one corner clipped \cite{hales, ToJi09b}. Another one involves the
Archimedean snub cube and snub dodecahedron, which are chiral
shapes (and therefore, non-centrally symmetric) \cite{Dijkstra}.
However, these chiral polyhedra still possess pairs of parallel
faces as a centrally symmetric particle, which allows the maximum
number of face-to-face contact when aligned. Nonetheless, these
special cases are overwhelmed in number by the set of those whose
optimal packings are not lattices. We also note that if Conjecture
2 is valid, it also applies to nonspherical particles derived by
smoothing the vertices, edges and faces of polyhedra provided that
the local curvature at face-to-face contacts is sufficiently
small.

Finally, we note that for certain centrally symmetric particles
that do not possess three equivalent principal axes (e.g.,
ellipsoids \cite{DonevEllip}), the rotational degrees of freedom
of the particle can be explored, resulting in dense
non-Bravais-lattice packings. This observation had led us to the
following conjecture \cite{ToJi10}:


\bigskip
\noindent{\bf Conjecture 3}: {\sl The densest packings of
congruent, centrally symmetric particles that do not possesses
three equivalent principle axes (e.g., ellipsoids) can be
non-Bravais lattices.}
\bigskip

\noindent {\it Remarks}: For a general smoothly-shaped particle,
the local principal curvature of the particle surface plays an
important role in determining their densest packing
configurations. This is to contrast polyhedral particles with flat
faces, whose dense packings are determined solely by maximizing
face-to-face contacts. For ellipsoids, the local principal
curvature at any point on the surface possesses non-trivial
values. Bezdek and Kuperberg were the first to show that
the densest packings of very elongated ellipsoids cannot be
Bravais lattice packings \cite{Bezkuper}. In particular,
by inserting very elongated ellipsoids into
cylindrical void channels passing through the ``ellipsoidal''
analogs of the face-centered-cubic sphere packing (affinely
deformed face-centered-cubic packings of spheres),
Bezdek and Kuperberg constructed congruent ellipsoid packings
whose density exceeds $\pi/\sqrt{18}$ and approaches 0.7459 in the limit
of infinitely thin prolate spheroids (i.e., the ``needle'' limit).
Later, Wills \cite{Wills} showed that similar constructions can be obtained
by inserting needle-like ellipsoids into
cylindrical void channels passing through an affinely
deformed hexagonal-close-packed lattice of spheres, leading to an improved
density of 0.7585. More recently, Donev et al. \cite{DonevEllip} found a family of
unusually dense non-Bravais-lattice packings of ellipsoids
with {\it arbitrary} aspect ratio. The packings in this family have
two ellipsoids per fundamental cell, rotated by $\pi/2$
relative to one another, and possess packing densities as high
as 0.7707 when the maximal aspect ratio is larger than $\sqrt{3}$,
which is close to the sphere shape.


Finally, we note that non-Bravais lattice
packings of elliptical cylinders (i.e., cylinders with an elliptical basal face)
that are denser than the corresponding optimal lattice packings
have been constructed \cite{Bezkuper}. This result supports our Conjecture 3.

\subsection{Organizing Principles for Concave Particles}

As pointed in the previous section, dense packings of convex
polyhedra favor a large number of face-to-face contacts between
neighboring particles, which can be achieved by either a lattice
packing or a periodic one depending on the particle
characteristics. It should not go unnoticed that our
arguments leading to our conjectures and propositions thus far do
not rely on the convexity of the particle shape and hence does not
seem to be a strongly limiting condition. Indeed, in this section,
we generalize the propositions to obtain appropriate propositions
for concave polyhedron particles.

We state the generalization of
Proposition 1 for concave particles as a proposition:

\bigskip
\noindent{\bf Proposition 3}: {\sl  Dense packings of centrally
symmetric concave, congruent polyhedra are given by their
corresponding densest lattice packings, providing a tight density
lower bound that may be optimal. }
\bigskip

\noindent Similarly, the generalization of Proposition 2 is given
by the following proposition:

\bigskip
\noindent{\bf Proposition 4}: {\sl Dense packings of concave,
congruent polyhedra without central symmetry are composed of
centrally symmetric compound units of the polyhedra with the
inversion-symmetric points lying on the densest lattice associated
with the compound units, providing a tight density lower bound
that may be optimal.}
\bigskip

\noindent {\it Remarks}: Due to the translational symmetry of a
periodic packing, the centrally symmetric compound unit can be
located anywhere within the fundamental cell. In practice, it is
usually more convenient to place the centroid of an original
polyhedron on the lattice points than to place the inversion
center of the compound unit on them. Note also that for concave
shapes, even the densest packing could possess a density that is
much smaller than unity. Therefore, the term ``dense'' in the
above two propositions means that the packing possesses a
relatively high density compared to that of other packings of the
same concave shape.

\begin{figure}[htbp]
$\begin{array}{c@{\hspace{1.0cm}}c}
\includegraphics[width=4.5cm,keepaspectratio]{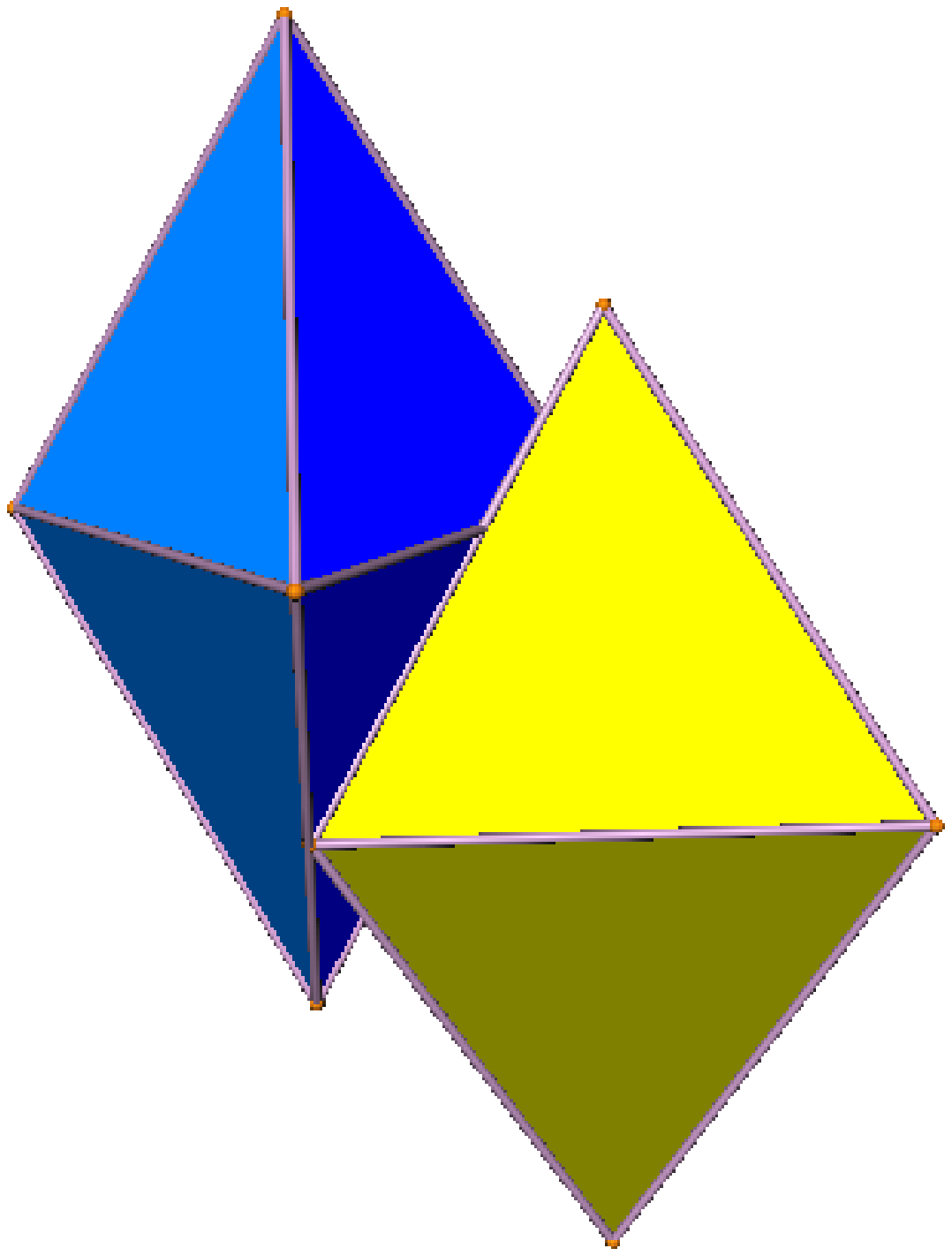} &
\includegraphics[width=4.5cm,keepaspectratio]{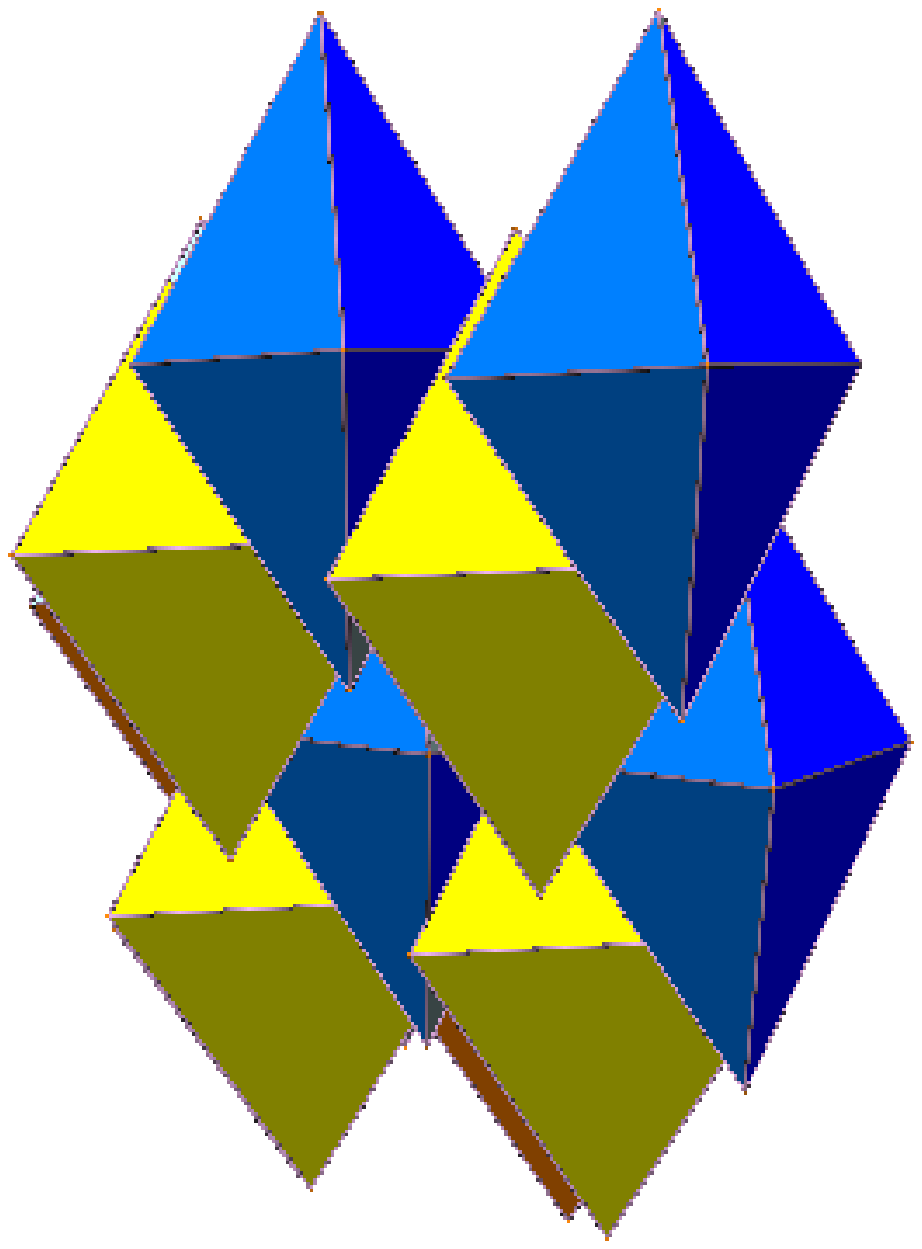} \\
{\mbox (a)} & {\mbox (b)} \\
\end{array}$
\caption{(Color online) (a) A centrally symmetric concave compound unit of
tetrahedra, which is composed of two contacting dimers of
tetrahedra. (b) A portion of the densest known packing of
tetrahedra, which is a lattice packing of the centrally symmetric
compound unit.} \label{concave_packing}
\end{figure}

Our new propositions for concave shapes are in fact motivated by
our investigation of the densest known packings of regular
tetrahedra, which are periodic packings with four particles in the
fundamental cell \cite{Ka10, ToJi10, Ch10}. It was already noted
in Ref. \cite{ToJi10} that the 4-particle compound object in the
fundamental cell composed of two contacting dimers (i.e., two
tetrahedra forming perfectly face-to-face contact) possesses
center inversion symmetry and thus, it can be viewed as a
centrally symmetric {\it concave polyhedron}, see
Fig.~\ref{concave_packing}. Hence, the densest known packings of
tetrahedra are in fact Bravais-lattice packings of such {\it
concave compound polyhedra}. Based on the same arguments leading
to Propositions 1 and 2 \cite{ToJi09, ToJi09b}, it is not
surprising that dense Bravais-lattice packings of the centrally
symmetric compound polyhedra have a high density, since there are
a large number of face-to-face contacts, which are made possible
due to the central symmetry of the compound polyhedron and bring
the centroids of the objects closer to each other, despite of the
concave nature of the compound polyhedron.


\begin{table}[htbp]
\label{tab_poly2} \caption{Characteristics of dense packings of
certain concave polyhedral particles, including whether the
particle possesses central symmetry, whether the densest known
packing is a Bravais lattice packing, the number of basis particles in the
fundamental cell if the packing is a non-Bravais-lattice periodic packing \cite{Dijkstra},
the numerically obtained density $\phi^*_{lat}$
for the lattice packing \cite{Dijkstra}, the highest numerically
obtained density $\phi^*$ \cite{Dijkstra}, and the upper bound on
the density $\phi^U_{max}$ \cite{Dijkstra}.}
\begin{tabular}{c@{\hspace{0.1cm}}c@{\hspace{0.1cm}}c@{\hspace{0.35cm}}c@{\hspace{0.35cm}}c@{\hspace{0.35cm}}c}
\hline\hline
Polyhedron & Cen. Sym. & Structure & $\phi^*_{lat}$ & $\phi^*$ & $\phi^U_{max}$ \\
\hline
Cs{\'a}sz{\'a}r Polyhedron & No & Periodic, 4-particle basis & 0.381 & 0.631 & 1 \\
Szilassi Polyhedron & No & Periodic, 2-particle basis & 0.331 & 0.519  & 1 \\
Echidnahedron & Yes & Bravais Lattice & 0.294 & 0.294 & 1 \\
Escher's Solid & Yes & Bravais Lattice & 0.947 & 0.947 & 1 \\
Great Rhombic Triacontahedron & Yes & Bravais Lattice & 0.557 & 0.557 & 1 \\
Jessen's Orthogonal Icosahedron & Yes & Bravais Lattice & 0.749 & 0.749 & 1 \\
Rhombic Dodecahedron Stellation II & Yes & Bravais Lattice & 0.599  & 0.599 & 1 \\
Great Stellated Dodecahedron & Yes & Periodic, 2-particle basis & 0.857  & 0.889 & 1 \\
Rhombic Hexecontahedron & Yes & Periodic, 2-particle basis & 0.529 & 0.556  & 1 \\
Small Triambic Icosahedron & Yes & Periodic, 2-particle basis & 0.688 & 0.695 & 0.977 \\
Nanostar & Yes & Bravais Lattice & 0.686 & 0.686 & 1 \\
Octapod & Yes & Bravais Lattice & 0.310 & 0.310 & 1 \\
Tetrapod & No & Periodic, 2-particle basis & 0.591 & 0.592 & 1 \\
\hline\hline
\end{tabular}
\end{table}

Very recently, de Graaf, van Roij and Dijkstra \cite{Dijkstra}
numerically generated the densest known packings of a variety of
concave particles, including certain well-known concave polyhedra
and approximations of real concave nanoparticles; see Fig. 8 for
examples of dense packings of certain concave nonspherical particles
reported in Ref. \cite{Dijkstra}. In Table~III, we
provide certain packing characteristics of a subset of the concave
polyhedra studied by these authors.

\begin{figure}[htbp]
$\begin{array}{c}
\includegraphics[width=8.0cm,keepaspectratio]{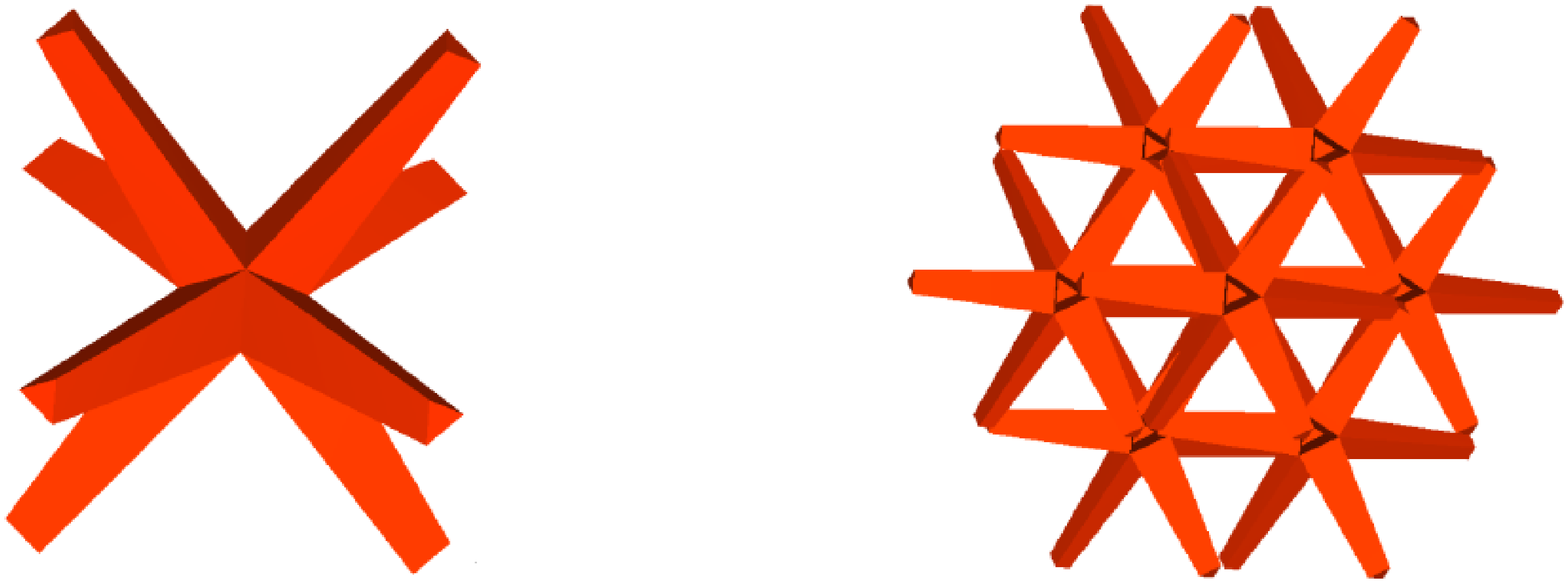} \\
\includegraphics[width=7.0cm,keepaspectratio]{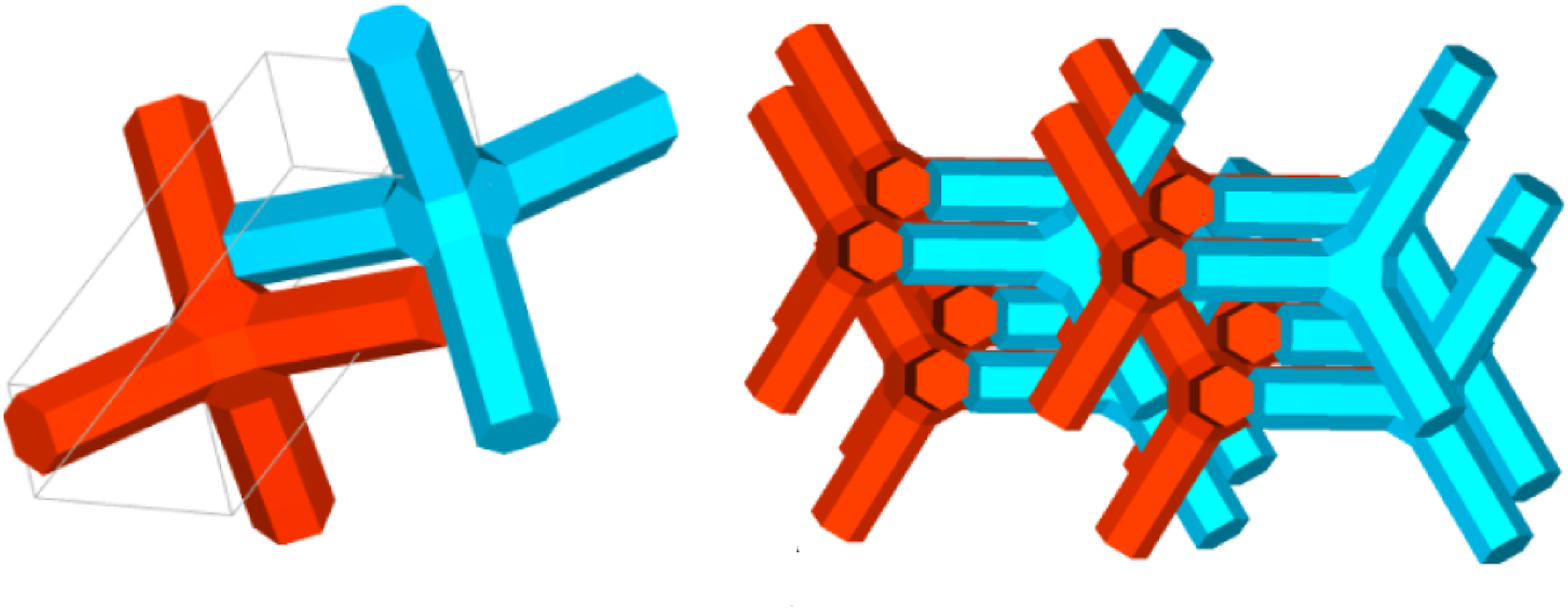} \\
\end{array}$
\caption{(Color online) Densest known packings of concave particles (adapted from
Ref.~\cite{Dijkstra}, image courtesy of de Graaf, van Roij and Dijkstra).
Upper panel: a centrally symmetric octapod (left)
and the corresponding densest known packing (right), which is a lattice
packing of octapods. Lower panel: a dimer of non-centrally symmetric tetrapods (left)
and the corresponding densest known packing (right), which is
periodic packing with two tetrapods in the fundamental cell.}
\label{concave_packingII}
\end{figure}

Despite the concavity of these particles, it can be clearly seen
from Table~III that for centrally symmetric shapes, the densest
lattice packings either correspond to the obtained densest
packings or provide a very tight lower bound on the densest
packings, which are consistent with our propositions. In fact, for
the three centrally symmetric solids (great stellated
dodecahedron, rhombic hexecontahedron, and small triambic
icosahedron) whose densest packings obtained are not lattice
packings, the difference in the packing density between the
obtained lattice packing and densest packing is very small (i.e.,
less than $3\%$). Given the complexity of the particle shapes
involved, it could be possible that the numerical method was not
able to obtain the corresponding optimal lattice packings. On the
other hand, for shapes lacking central symmetry, the densest
packings always correspond to lattice packings of the centrally
symmetric compound objects composed of the original polyhedra, and
thus, are periodic packings of the original polyhedra. We note
that due to the large asphericity (i.e., the ratio of circumsphere
radius over insphere radius of the particle \cite{ToJi09b})
associated with concave shapes, the upper bound (\ref{lemma}) for
many concave polyhedra gives trivial value of unity, which is
usually much larger than the highest packing densities obtained.
Nonetheless, the robustness of the numerical procedure for
generating the dense packings of concave particles has been
verified by virtually producing the densest known packings of the
Platonic and Archimedean solids, among other shapes
\cite{Dijkstra}. Therefore, it is reasonable to believe that most
of the numerical packings of the concave particles obtained from
the computer simulations in Ref.~\cite{Dijkstra} are close to
optimal. Nonetheless, all of the densest known packings of the
large set of the concave particles shown in Table III are
consistent with Propositions 3 and 4.


\begin{figure}[htbp]
$\begin{array}{c@{\hspace{1.0cm}}c}
\includegraphics[width=3.5cm,keepaspectratio]{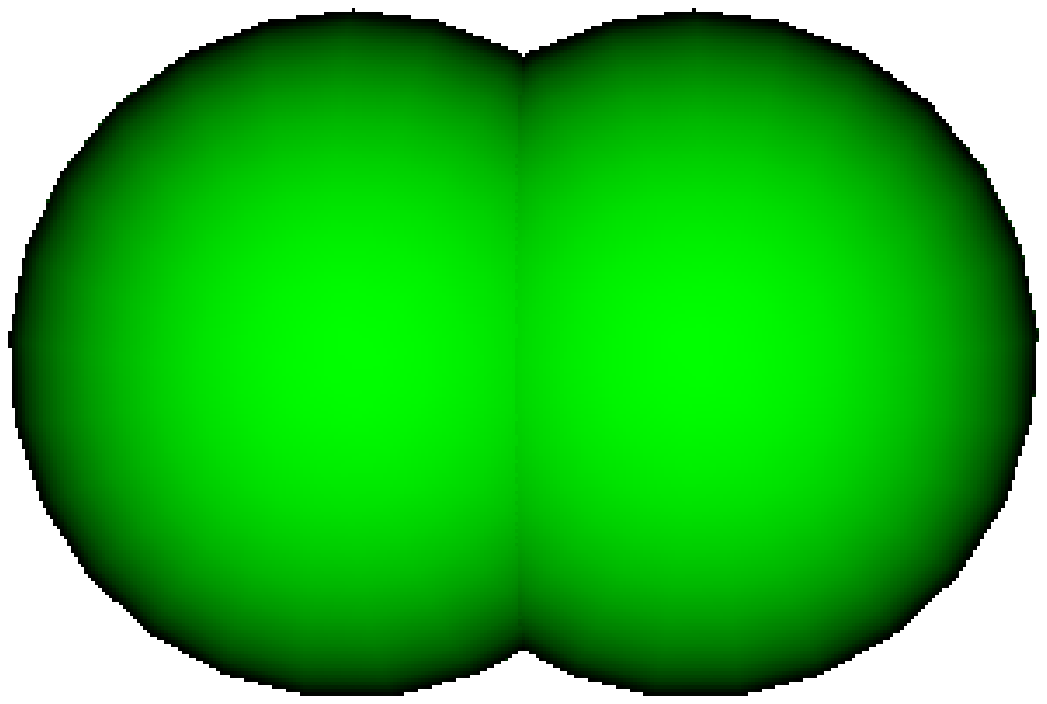} &
\includegraphics[width=3.5cm,keepaspectratio]{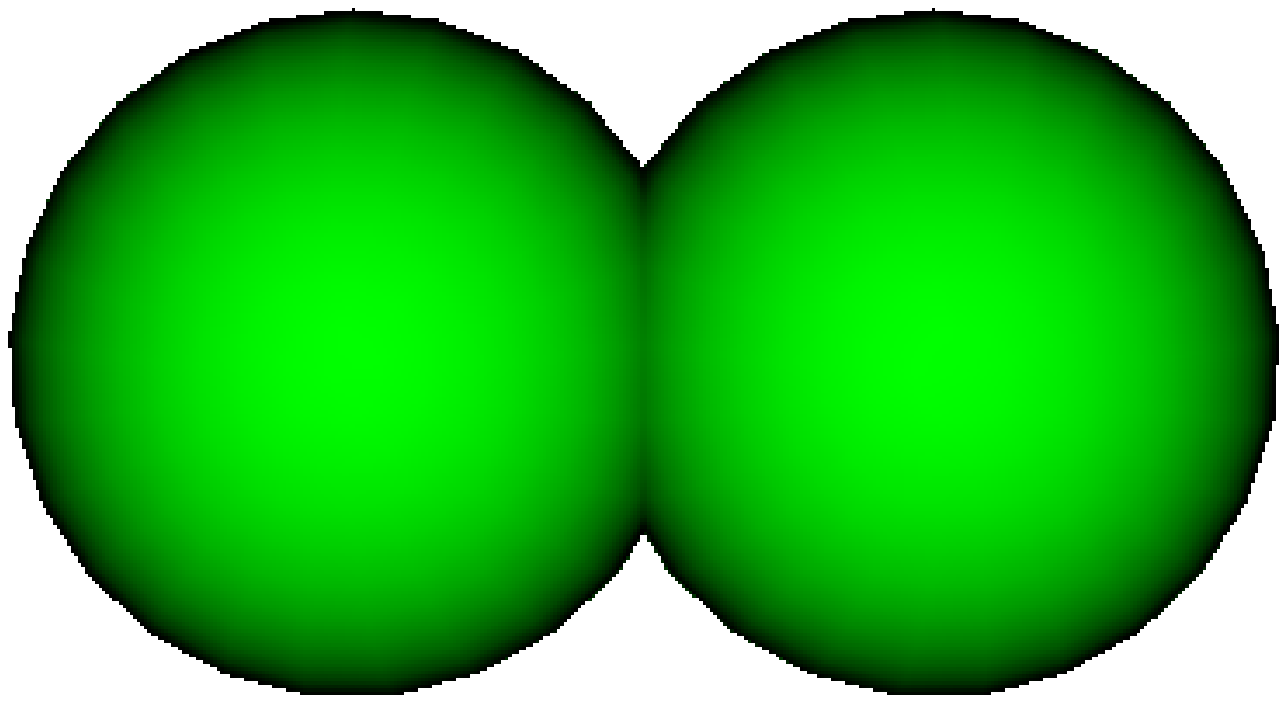} \\
\end{array}$
\caption{(Color online) Dimers of spheres of radius $R$, with sphere center
separation $L$. Left panel: A dimer of spheres with aspect ratio
$L/R = 0.85$. Right panel: A dimer of spheres with aspect ratio
$L/R = 1.85$, in which the two spheres just touch each other.}
\label{fig_sphere_dimer}
\end{figure}

Finally, we note that our Propositions 3 and 4 should also apply
for certain smoothly-shaped concave particles. For example,
consider the family of dimers formed by two spheres in
$\mathbb{R}^3$ that can overlap to various degrees (see
Fig.~\ref{fig_sphere_dimer}), whose volume is given by
\begin{equation}
V_0 = \frac{4\pi}{3}R^3 \left ( {1+\frac{3}{4}\frac{L}{R}-\frac{1}{16}\frac{L^3}{R^3}} \right),
\end{equation}
where $R$ is the radius of the spheres and $L \in [0, 2R]$ is the separation between the sphere
centers. This family of dimers of spheres includes the single-sphere case
and the two-touching-sphere case as extreme cases. Since such a
dimer possesses central symmetry, the densest packing of this
shape can be expected to be a lattice packing according to
Propositions 3. Indeed, it has been shown that in both
$\mathbb{R}^2$ \cite{2d_dimer} and $\mathbb{R}^3$ \cite{3d_dimer},
the densest known dimer-sphere packings are achieved by certain
Bravais lattices.


\section{Additional Applications of the Organizing Principles}

In this section, we illustrate how the general organizing
principles formulated in the previous sections can be applied in
practice. Specifically, Propositions 1-4 provide constructable
tight lower bounds on dense packings of both centrally symmetric
polyhedra and the ones lacking central symmetry. These organizing
principles should enable one to construct dense packings of a
specific nonspherical shape, which are either optimal or close to
optimal. We demonstrate this capability by finding the densest
known packings of centrally symmetric spherocylinders, and
``lens-shaped'' particles as well as non-centrally symmetric
square pyramids and rhombic pyramids. In addition, we show that
analytically constructed dense packing configurations as
determined by our organizing principles can serve as the starting
point to study the equilibrium phase behavior of the corresponding
particles at high densities \cite{TrunTetrah, Aga11, Bob,
GlotzerPhase, ellip_II, ellip_III, ellipPhase}.

\subsection{Dense packings of spherocylinders}

\begin{figure}[htbp]
\begin{center}
$\begin{array}{c}
\includegraphics[width=7.5cm,keepaspectratio]{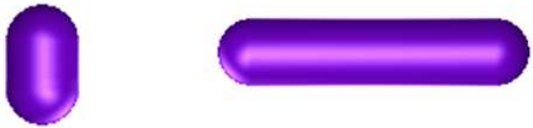} \\
\end{array}$
\end{center}
\caption{(Color online) Spherocylinders composed of cylinder with length $L$,
caped at both ends with hemispheres with radius $R$. Left panel: A
spherocylinder with aspect ratio $L/R = 1$. Right panel: A
spherocylinder with aspect ratio $L/R = 5$.}
\label{fig_spherocylinder}
\end{figure}

Let us now consider the case of a spherocylinder. A spherocylinder
consists of an cylinder with length $L$ and
radius $R$ capped at both ends by hemispheres with radius $R$ (see
Fig. \ref{fig_spherocylinder}), and therefore is a centrally-symmetric
convex particle. Its volume is given by
\begin{equation}
V_0 = \pi R^2 L + \frac{4}{3}\pi R^3.
\end{equation}
At the limit $L=0$, a spherocylinder reduces to a sphere with radius $R$.

For spherocylinders with $L>0$, it appears that the
densest known Bravais-lattice packing could be the actual densest
packing of the shape. This is because the local principal
curvature of the cylindrical surface is zero along the spherocylinder axis,
and thus, spherocylinders can have very dense lattice packings by
aligning the cylinders along their axes.
The density of the optimal Bravais-lattice packing of spherocylinders
is given by
\begin{equation}
\label{sphero_phi}
\phi =\frac{\pi}{\sqrt{12}}~\frac{L+\frac{4}{3}R}{L+\frac{2\sqrt{6}}{3}R},
\end{equation}
where $L$ is the length of the cylinder and $R$ is the radius of
the spherical caps \cite{fn_sc_packing}.
However, we note that there are a noncountably-infinite
number of non-Bravais-lattice packings of spherocylinders with the
same packing densities (\ref{sphero_phi}),
associated with the infinite number of different stackings of spherocylinder layers
similar to the Barlow stackings of the spheres \cite{ToStRMP}. Thus,
the set of dense non-lattice packings of spherocylinders is overwhelmingly
larger than that of the lattice packing. We emphasize that the
role of the local principal curvature of the particle surface in
determining dense packing configurations is far from being
completely understood.

\subsection{Dense packings of ``lens-shaped'' particles}

\begin{figure}[htbp]
\begin{center}
$\begin{array}{c}
\includegraphics[width=7.5cm,keepaspectratio]{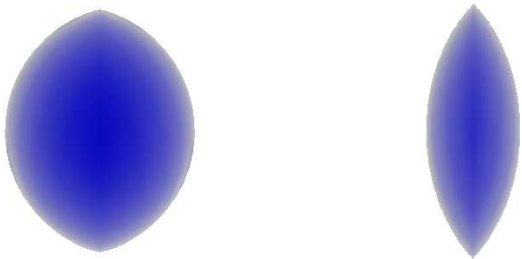} \\
\end{array}$
\end{center}
\caption{(Color online) ``Lens-shaped'' particles corresponding to the intersection
regions of two spheres with radius $R$ separated by distance $L$.
Left panel: A lens-shaped particle associated with $L/R = 0.8$. Right panel: A
lens-shaped particle associated with $L/R = 1.6$.}
\label{fig_lens}
\end{figure}

Let us also consider dense packings of ``lens-shaped'' particles, which
correspond to the intersection region of two congruent spheres,
and therefore is a centrally-symmetric convex particle.
In particular, the volume of such a lens-shaped particle is given by
\begin{equation}
V_0 = \frac{4\pi}{3}R^3 \left ( {1-\frac{3}{4}\frac{L}{R}+\frac{1}{16}\frac{L^3}{R^3}} \right),
\end{equation}
where $R$ is the radius of the spheres and $L \in [0, 2R)$ is the
separation distance between the sphere centers.

Similar to the aforementioned case of spherocylinders, it appears
that both the optimal Bravais-lattice packing of such lens-shaped
particles, corresponding to the FCC-like stacking of the
triangular-lattice layers of the particles, and the
noncountably-infinite number of Barlow-like stackings of the
layers give the densest known packings, which possess densities
given by
\begin{equation}
\phi = \frac{8\pi}{\sqrt{3}} \frac{1-\frac{3}{4}\frac{L}{R}+\frac{1}{16}\frac{L^3}{R^3}}
{(4-\frac{L^2}{R^2})[(24+\frac{3L^2}{R^2})^{\frac{1}{2}}-\frac{3L}{R}]}.
\end{equation}
Note that in the sphere limit (i.e., $L=0$), one can obtain the
optimal sphere-packing density $\phi = \pi/\sqrt{18}$. In the
limit of infinitely thin ``lenses'' ($L\rightarrow 2$), one has
$\phi = \sqrt{3}\pi/8$. However, we do not exclude the possibility
that there exist other Bravais-lattice or non-lattice packings
that are denser than our current constructions.


\subsection{Dense packings of pyramids from octahedron packing}

\begin{figure}[htbp]
\begin{center}
$\begin{array}{c@{\hspace{1.0cm}}c@{\hspace{1.0cm}}c}
\includegraphics[width=4.0cm,keepaspectratio]{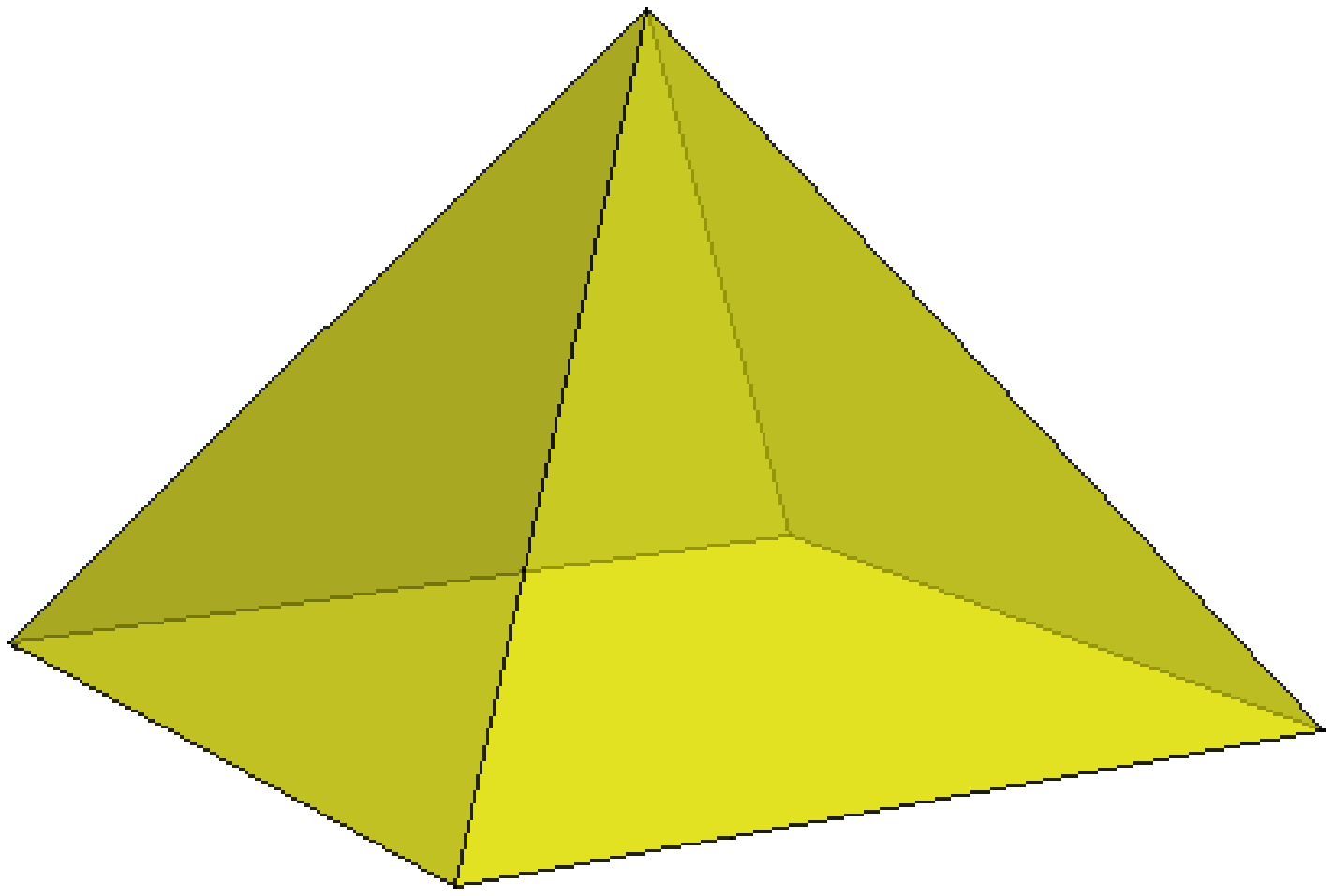} &
\includegraphics[width=4.0cm,keepaspectratio]{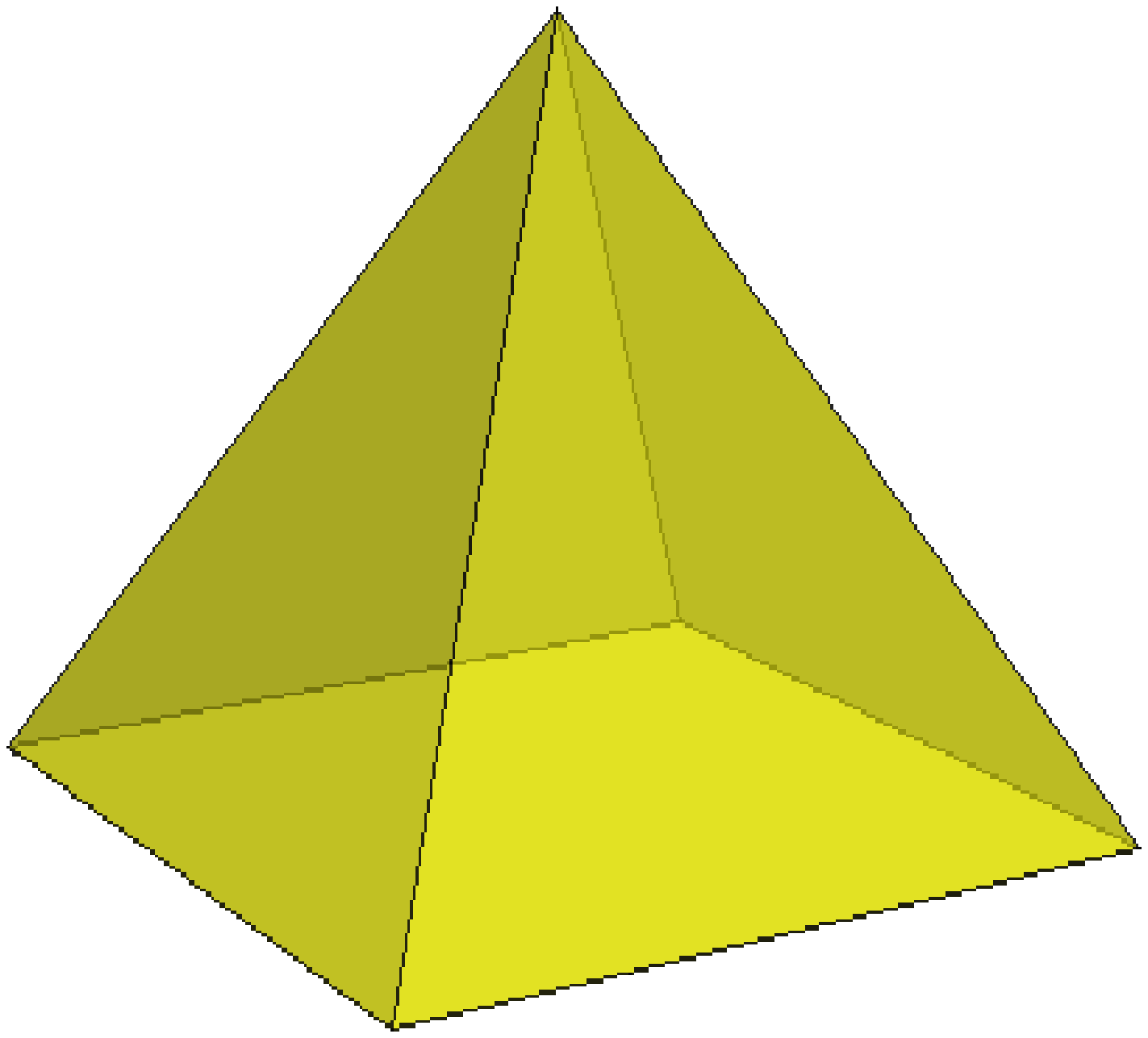} &
\includegraphics[width=4.0cm,keepaspectratio]{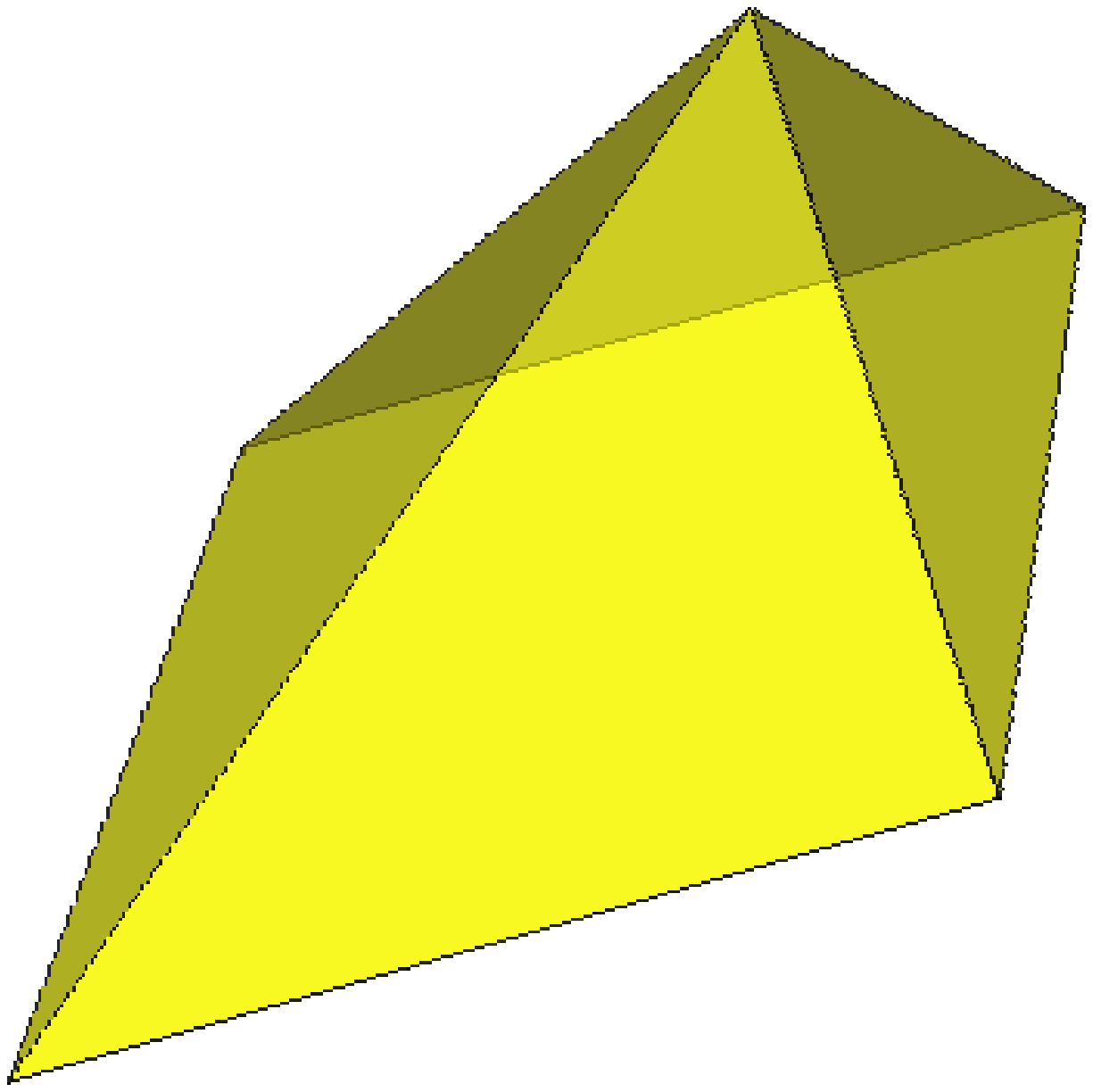} \\
{\mbox (a)} & {\mbox (b)} & {\mbox (c)} \\
\end{array}$
\end{center}
\caption{(Color online) (a) A square pyramid with $\ell = \sqrt{2}h$. (b) A
stretched square pyramid with $\ell < \sqrt{2}h$. (c) A sheared
pyramid with a rhombic basal plane. } \label{fig_pyramid_shape}
\end{figure}

\begin{figure}[htbp]
\begin{center}
$\begin{array}{c@{\hspace{1.0cm}}c@{\hspace{1.0cm}}c}
\includegraphics[width=4.0cm,keepaspectratio]{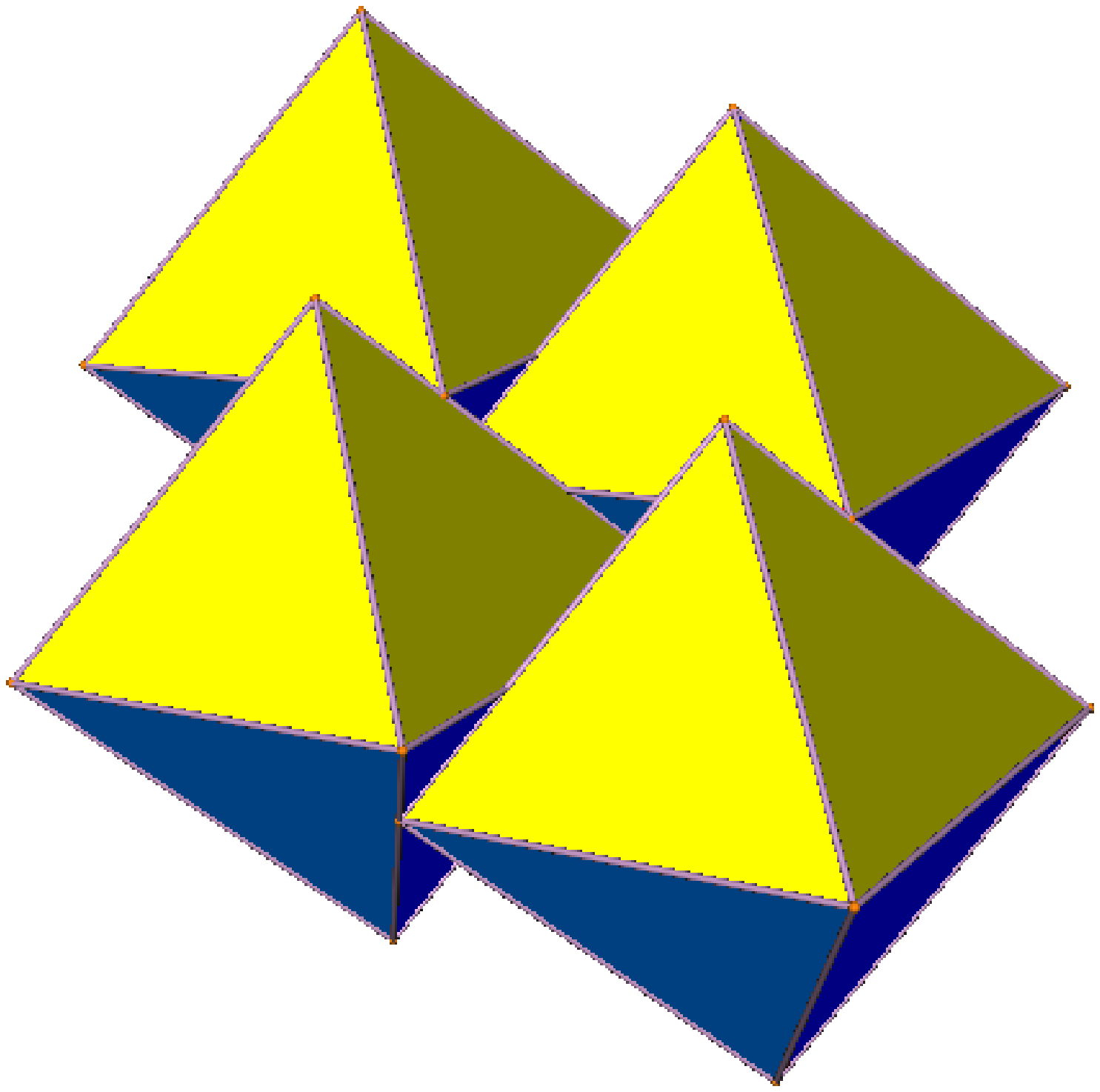} &
\includegraphics[width=4.0cm,keepaspectratio]{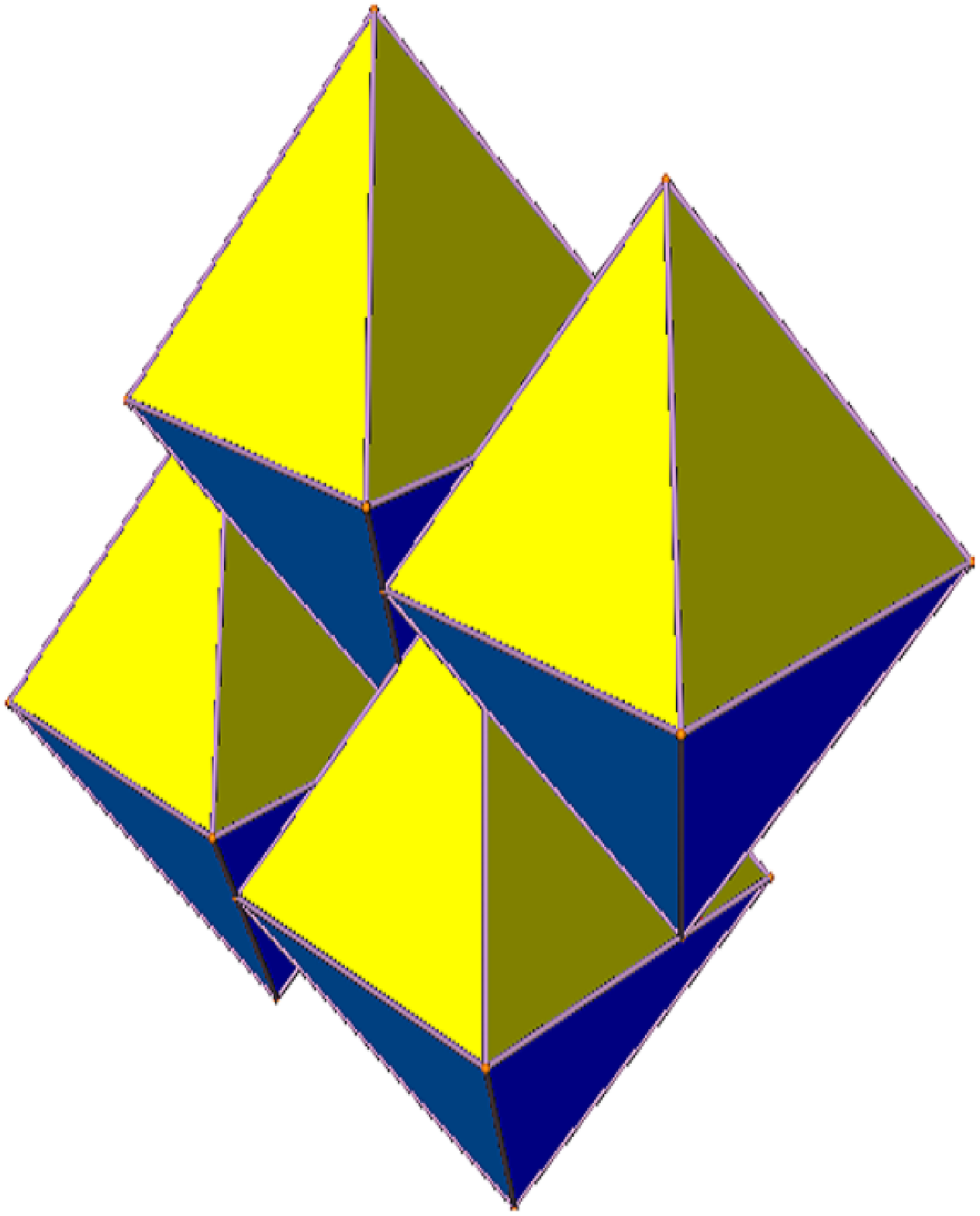} &
\includegraphics[width=4.0cm,keepaspectratio]{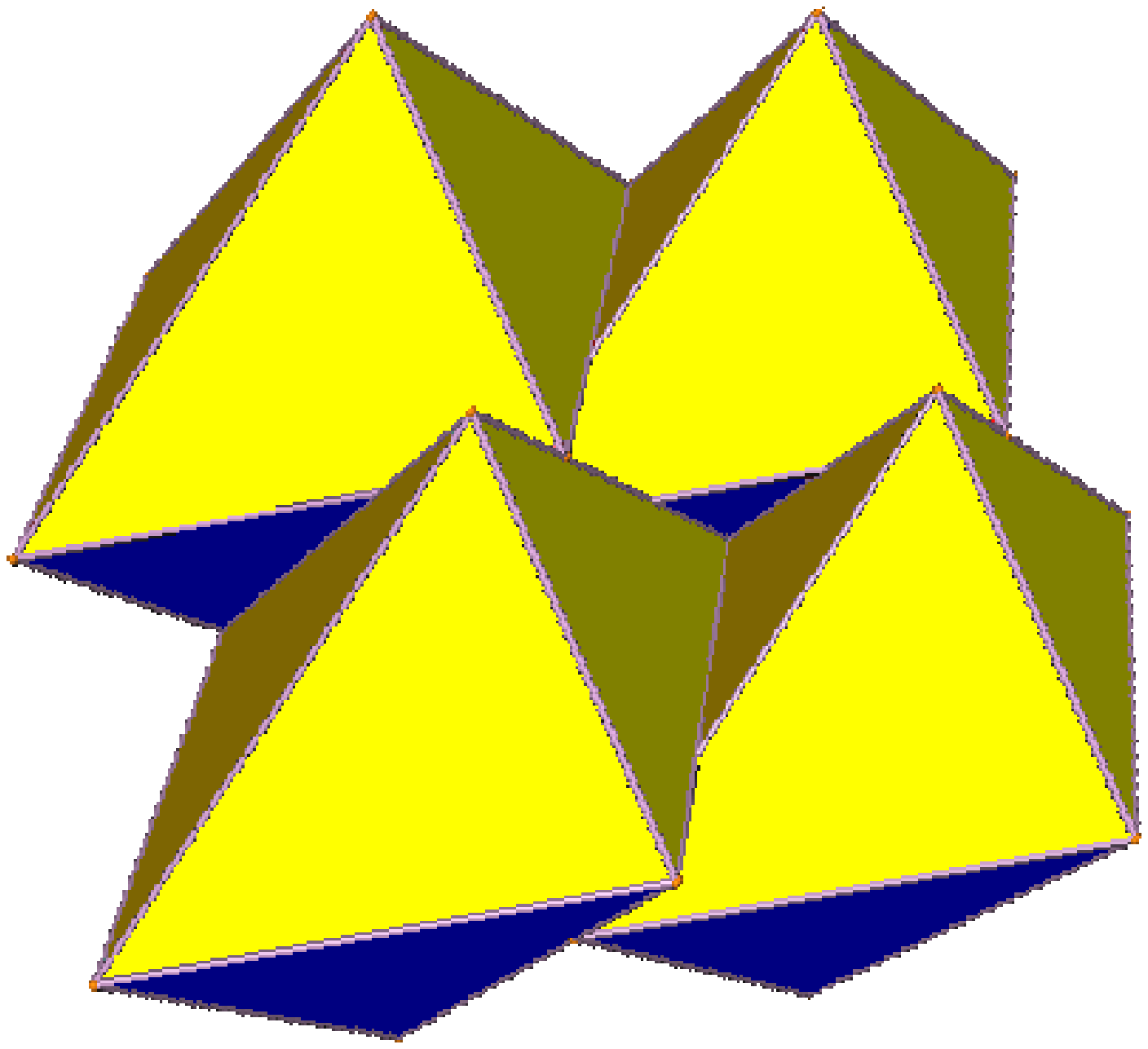} \\
{\mbox (a)} & {\mbox (b)} & {\mbox (c)} \\
\end{array}$
\end{center}
\caption{(Color online) (a) A portion of a dense periodic packing of the square
pyramids with $\ell = \sqrt{2}h$, whose fundamental cell contains
two pyramid forming a centrally symmetric octahedra. (b) A portion
of a dense periodic packing of the stretched square pyramids with
$\ell < \sqrt{2}h$, which corresponds to an affine transformation
of the densest packing of octahedra. (c) A portion of a dense
periodic packing of the sheared pyramids, which corresponds to a
sheared packing of octahedra. } \label{fig_pyramid}
\end{figure}

Consider a square pyramid with height $h$ and length $\ell$ of the
basal square (see Fig.~\ref{fig_pyramid_shape}). According to
Proposition 2, the square pyramid lacks central symmetry and dense
packings of this shape should be lattice packings of a centrally
symmetric compound object made of the pyramids. We first consider
the case when $\ell = \sqrt{2}h$, so that a pair of the pyramids
that form a perfect contact through the square faces correspond to
the regular octahedron (see Fig.~\ref{fig_pyramid}a). Therefore, a
dense packing of this special pyramid can be immediately obtained,
which is exactly the densest lattice packing of octahedra with
density $\phi = 18/19 = 0.974368\ldots$.

For the square pyramid with $\ell \neq \sqrt{2}h$, a centrally
symmetric compound dimer can still be constructed in a similar
way, which corresponds to a octahedron that is stretched ($\ell <
\sqrt{2}h$) or compressed ($\ell > \sqrt{2}h$) along the direction
of the height of the pyramids. To obtain dense packings of such
pyramids, one can apply an affine transformation to the densest
lattice packing of regular octahedron along one of its principal
directions (see Fig.~\ref{fig_pyramid}b), which leaves the packing
density unchanged. This leads to a dense packing of stretched
(compressed) square pyramids with $\phi = 18/19 = 0.974368\ldots$
Similarly, for pyramids whose basal plane is a rhombus obtained by
uniformly shearing the square, one can construct a dense packing
of such objects by uniformly shearing the densest regular
octahedron packing in the plane defined by a pair of the principal
directions of an octahedra, resulting in the same density (see
Fig.~\ref{fig_pyramid}c).

We note that the constructed packings of the various
aforementioned pyramids may not correspond to their associated
densest packings, but these constructions should be close to the
optimal ones. Once a construction is obtained by applying the
organizing principles, one can perform various types of
optimizations to improve the density by finding the (locally)
densest configuration \cite{ToJi10}.

\subsection{Equilibrium High-Density Crystal Phases}

At infinite pressure, the optimal (densest) packing of hard
particles corresponds to the thermodynamic equilibrium phase for
these particles. In the case of identical hard spheres, the
face-centered-cubic lattice is the thermodynamically favored
maximal-density state; see Fig. 2. Colloidal particles can
synthesized to interact with effective hard-particle potentials
\cite{ToStRMP}, and thus, it is useful to see to what extent the
solid phase associated with the densest packing structure is
stable under finite pressure by decompressing the packing using
computer simulations \cite{Frenkel_Book}. Therefore, our general
organizing principles enable one to obtain the equilibrium
crystalline phase behavior of the hard particles at high
densities. To illustrate this application, we consider the
decompression of the densest known packing of truncated
tetrahedra.

\begin{figure}[htbp]
\begin{center}
$\begin{array}{c}
\includegraphics[width=12.5cm,keepaspectratio]{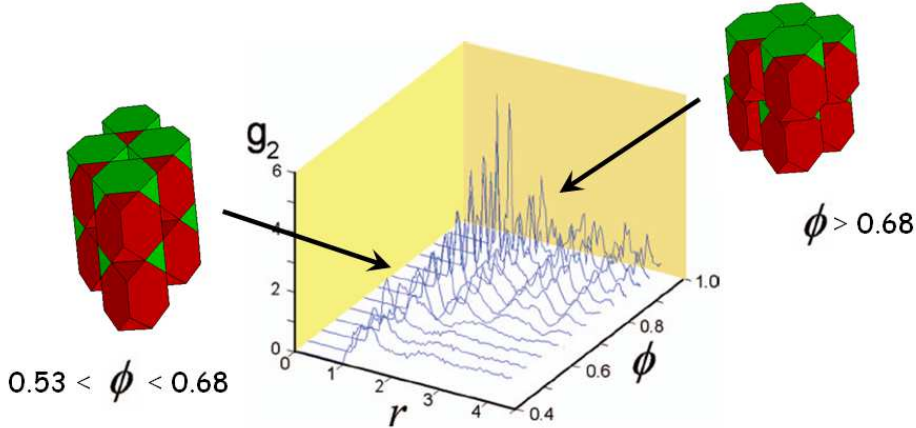} \\
\end{array}$
\end{center}
\caption{(Color online) Pair correlation function $g_2(r)$
associated with the centroids of the truncated tetrahedra at
different densities in the range from 0.99 to 0.38 during the
melting of the highest-packing-fraction crystal \cite{TrunTetrah}.
Our simulation suggests that the melting process has two
stages, including a solid-solid transition at a high density and a
solid-liquid transition at an intermediate density
\cite{TrunTetrah}. At very high densities, $g_2(r)$ is a
long-ranged function and the stable phase is associated with the
densest packing of truncated tetrahedra. At intermediate
densities, the stable phase is associated with a less dense
packing of truncated tetrahedra with higher symmetry. Below $\phi
\approx 0.53$, $g_2(r)$ suddenly changes from a long-ranged
function to an exponential decaying function, indicating the
occurrence of a first-order crystal-liquid transition.}
\label{fig_g2}
\end{figure}

For truncated tetrahedra, Monte-Carlo (MC) simulations have been
carried out \cite{TrunTetrah} to ``melt'' the densest known
packing structure via a decompression process. In particular, a
periodic simulation box containing $N=686$ particles is employed,
whose size and shape are allowed to change \cite{ToJi09, ToJi09b}.
The volume of the simulation box is slowly increased to decrease
the pressure and density (i.e., packing fraction) of the system.
At each density, a total number of 10 million MC trial moves are
applied to each particle and $100~000$ trial volume-preserving
deformations are applied to the simulation box to equilibrate the
systems. Equilibrium structural characteristics, such as the
pair-correlation function $g_2$ (see Fig.~\ref{fig_g2}) and the
number of dimers $n_2$ of truncated tetrahedra, are collected.
Such descriptors are used to gauge the remaining crystalline order
in the system during the decompression process. It is found that
above $\phi \approx 0.68$, the crystal configurations associated
with the optimal packing of truncated tetrahedra is the stable
solid phase for these particles. When $0.53 < \phi < 0.68$, the
crystal configurations associated with a packing with lower
density but higher symmetry \cite{TrunTetrah} becomes the stable
phase. This suggests that the melting process has two stages,
including a solid-solid transition a high density and a
solid-liquid transition at an intermediate density
\cite{TrunTetrah}. Whether the crystalline phases are indeed the
ones that we have identified via the decompression process must
still be verified using other techniques \cite{fn_melting}. Below
$\phi \approx 0.53$, the correlation function $g_2$ suddenly
changes from a long-ranged function to an exponential decaying
function (see Fig.~\ref{fig_g2}) and $n_2$ quickly drops from
$N/2$ to almost zero, indicating the occurrence of a first-order
crystal-liquid transition.

The wide range of stability for the crystal phases [$\phi \in
(0.53, 0.995)$] is due to the fact that the dimers of truncated
tetrahedra can fill space very efficiently, as can be seen from
the amazingly high packing density (i.e., $207/208 = 0.995\ldots$;
see Table I). This means that the free volume associated with
crystals of the truncated tetrahedra is readily maximized in the
dimer arrangement, leading to a lower free energy of the system.
Since a dimer is formed by a pair of truncated tetrahedra
contacting through the a common large hexagonal face, it is
relative easy for such local clusters to form in a relatively
dense liquid. Once such dimers nucleate, the system is expected to
crystallize easily upon further compression. We note that the
aforementioned decompression simulations can only provide
estimates of the phase transition densities, the exact coexistence
range of $\phi$ and whether there are higher-order solid-solid
phase transitions can be precisely explored by carrying out
free-energy calculations.

Similar complex high-density phase behaviors have also been
observed in other nonspherical-hard-particle systems. For example,
for ellipsoids with revolution at intermediate and large aspect
ratios, it has been shown from decompression simulations
\cite{ellip_II}, free-energy calculations \cite{ellip_III} and
replica exchange Monte Carlo simulations
\cite{ellipPhase} that the densest known 2-particle basis packings
constructed by Donev et al. \cite{DonevEllip} correspond to the
stable high-density equilibrium phases. As the density decreases,
nematic phases becomes the stable phases and the system undergoes
a solid-liquid-crystal phase transition \cite{ellip_II,
ellipPhase}. For cube-like superballs, decompressing the densest
known crystalline packings via molecular dynamics reveals the
stability of the crystal phases associated with these packings and
a solid-liquid-crystal transition to cubatic phases \cite{Bob}.
More recent free-energy calculations \cite{NiSuperball} also
showed that the densest known packings of superballs correspond to
the high-density equilibrium phases, while at intermediate
densities, instead of cubatic phases, the FCC-plastic-crystal
phases are stable.

The aforementioned complex melting processes associated with
nonspherical particles are much richer than for hard-sphere system
in which there is one type of crystal all the way down from the
densest packing to the melting point. Because the rotational
degrees of freedom and symmetries of a hard nonspherical particle
can result in a variety of different high-density phases (as
opposed to the simple case of identical spheres) \cite{TrunTetrah,
Aga11, Bob, GlotzerPhase, ellipPhase}, one must be careful to
extend our conjectures and propositions that apply to high
densities to predict equilibrium phases at intermediate densities
(e.g., liquid-solid transitions).

\section{Conclusions and Discussion}

In this paper, we have provided general organizing principles to guide one to obtain
maximally dense packings of nonspherical hard particles, and tested them against
the most comprehensive set of packings of both convex and concave
particles examined to date. For polyhedron particles, a general rule for achieving
high packing densities is form a large number of face-to-face
contacts, allowing particle centroids to be closer to one another.
Therefore, despite whether the particle is convex or concave,
central symmetry plays an important role in determining the dense
packing structures. For centrally symmetric (convex or
concave) particles, the densities of their densest Bravais-lattice
packings provide tight lower bounds on the corresponding maximal
packing densities that, according to Propositions 1 and 3, could
be the optimal ones. This is due to the fact that when all the
particles are aligned in the same direction, the central symmetry
of the particle allows a large number of face-to-face contacts.
For particles lacking central symmetry, their densest lattice
packing contains a large number of vertex-to-face contacts which
always leads to a low packing density. Instead, the densest
lattice packing of a centrally symmetric compound unit of the
original particles (i.e., a periodic packing of the original
particles) provides a tight lower bound on the maximal packing
density that, according to Propositions 2 and 4, could be the
optimal one.

Organizing principles for certain smoothly-shaped particles, such
as ellipsoids and superballs, have also been provided. A major
difference between such smooth particles and a polyhedron is that
the former possesses nontrivial surface curvature. A rule to
obtain dense packings of particles with smooth surfaces, similar
to that for polyhedra, is that the particles form contacts at
surface regions with small local principal curvature (i.e.,
relatively ``flat'' regions). Thus, for centrally symmetric
superballs, both numerical simulations and theoretical analysis
suggest that the densest lattice packing of these shapes would be
optimal. However, central symmetry alone does not seem to be
sufficient to ensure that the densest packing of smooth particles
is given by the corresponding optimal lattice packing. For
example, for ellipsoids, which do not possess equivalent principal
axes, periodic packings are constructed that are denser
than the optimal lattice packings. It appears that the elongated shape
of an ellipsoid results in local principal curvatures that
allows the rotational degrees of freedom of the particles to be
fully explored. As noted in Sec. III, the role of curvature in
determining dense packings for smoothly-shaped particles is still
not very well understood.

Moreover, we have shown how our generalized organizing
principles can be applied to construct analytically the densest
known packings of certain convex nonspherical particles (e.g.,
centrally symmetric spherocylinders and ``lens-shaped'', as well
as non-centrally symmetric square pyramids and rhombic pyramids).
We have also shown that these principles can be profitably applied
to infer the high-density crystalline phases of hard convex and
concave particles.

We emphasize that while the propositions that we have put forth 
concerning dense packings of congruent nonspherical packings in the 
present paper currently lack the rigor of strict mathematical proofs, 
we expect that they will either ultimately lead to strict proofs or at least 
provide crucial guidance in obtaining such proofs for some nonspherical
shapes. For example, in Ref. \cite{ToJi09b}, we provided
what we believe to be the major elements of a proof of Conjecture 1.


Can general principles be established that can be used to
categorize and characterize disordered jammed packings according
to the particle shapes and symmetries? This is in general a
notoriously difficult question to answer, since disordered
packings are intrinsically nonequilibrium states and can possess a
wide spectrum of densities and degrees of disorder \cite{ToStRMP,
JiaoJAP11}. Nonetheless, certain universal global structural
features have begun to emerge from recent studies of the maximally
random jammed (MRJ) states (analogs of the MRJ state for spheres
depicted in Fig. 1) of a wide class of nonspherical particles,
regardless of the particle shapes or relative sizes (see Table
IV). It has been shown that the MRJ packings possess special
particle number density or local volume fraction fluctuations
that vanish at large length scales
\cite{PlatonicMRJ, ChaseMRJ}, implying ``hyperuniformity'' of the
packings (vanishing of large-scale fluctuations) and leading to anomalous
quasi-long-range (QLR) correlations. This ``hyperuniformity''
property can be experimentally measured using standard scattering
techniques and is manifested as a linear small-wavenumber (i.e.,
small-$k$) behavior in the structure factor
\cite{hyperuniform_sal} or the power spectrum
\cite{hyperuniform_chase}. It has been argued that the
hyperuniform QLR correlations in MRJ packings arise from the
competition between the requirement of jamming and maximal
disorder. Therefore, the hyperuniform QLR correlations should be
expected for a wide class of MRJ packings.


\begin{table}[htp]
\centering \caption{Characteristics of MRJ packings of hard
particles with different shapes, including spheres \cite{ToStRMP},
ellipsoids \cite{DoSiSaVa04}, superballs \cite{JiStTo10},
superellipsoids \cite{DeCl10} and the non-tiling Platonic solids
\cite{PlatonicMRJ}.}
\begin{tabular}{c@{\hspace{0.35cm}}c@{\hspace{0.35cm}}c@{\hspace{0.35cm}}c}
\hline\hline
Particle & Isostatic & Hyperuniform QLR & MRJ Packing Fraction\\
\hline
Sphere &  Yes & Yes & 0.642 \\
Ellipsoid &  No (hypostatic) & Yes & $0.642-0.720$  \\
Superball &  No (hypostatic) & Yes & $0.642-0.674$ \\
Superellipsoid &  No (hypostatic) & Yes & $0.642-0.758$ \\
Octahedron &  Yes & Yes & $0.697$ \\
Icosahedron &  Yes & Yes & $0.707$ \\
Dodecahedron &  Yes & Yes & $0.716$ \\
Tetrahedron &  Yes & Yes & $0.763 $ \\
\hline\hline
\end{tabular}
\label{tab_MRJ}
\end{table}

In addition, recent studies of MRJ packings of polyhedra
\cite{Paul10, Fisher11, PlatonicMRJ} suggest that the iso-counting
conjecture proposed for sphere packings (i.e., each particle in
the packing possesses 2$n_{dof}$ constraints, where $n_{dof}$ is
the number of degrees of freedom) is also true for polyhedra (see
Table IV). We note that for polyhedra, the number of constraints
at each contact can be exactly given \cite{Paul10, Fisher11,
PlatonicMRJ}. On the other hand, for smoothly-shaped particles, it
was found that the MRJ packings are generally hypostatic, with
each particle possessing a smaller number of contacts than
2$n_{dof}$ \cite{DoSiSaVa04, Ohern09, JiStTo10, DeCl10,
spherocylinder}; see Table IV. Moreover, it was shown that the
surface curvature at contact points play an important role in
blocking the rotation of the particles and thus, in jamming the
packing. However, it is difficult to quantify the effective number
of constraints provided by each contact with different local
principal curvatures. A successful quantification of this
effective counting of constraints could lead to the conclusion
that the iso-counting conjecture also holds for smoothly-shaped
nonspherical particles.

Finally, we note that tunability capability via particle shape
provides a means to design novel crystal, liquid and glassy states
of matter that are much richer than can be achieved with spheres
\cite{ToStRMP, SalAPSTalk}. For example, by introducing particle asphericity,
a variety of optimal crystalline packings \cite{DonevEllip,
JiaoSuperdisk, JiaoSuperball, NiSuperball, Be00, Co06, ToJi09,
ToJi09b, tetrah2, Ka10, ToJi10, Ch10, kallus11, tiling,
TrunTetrah, GlozterTrunTetrah, Laura09, Dijkstra},
diverse disordered jammed packings \cite{DoSiSaVa04, Ohern09, JiStTo10, DeCl10, tetrah1,
spherocylinder, particulogy, tetrah3, Paul10, Fisher10, Fisher11,
PlatonicMRJ, ChaseMRJ} as well as a
wide spectrum of equilibrium phases \cite{TrunTetrah, Aga11, Bob, GlotzerPhase,
ellip_II, ellip_III, ellipPhase, GlotzerPhase2} have been obtained from computer simulations.
On the experimental side, it has been recently shown \cite{NatMat12} that silver 
polyhedral nanoparticles (with central symmetry) can self assemble into our 
conjectured densest lattice packings of such shapes \cite{ToJi09, ToJi09b}.
The resulting large-scale crystalline packings may facilitate the design and fabrication of
novel three-dimensional materials for sensing \cite{sensing},
nanophotonics \cite{nanophoto} and photocatalysis \cite{nanocata}.
Our general organizing principles can clearly provide valuable
guidance for novel materials design, which we will systematically
investigate in future work.  It will also be very useful to 
determine order maps for packings of nonspherical 
particles that are the analogs of Fig. 1 for sphere packings.

\begin{acknowledgments}
We are very grateful to Marjolein Dijkstra for giving us
permission to adapt the figures of the octapod and tetrapod that
appeared in Ref.~\cite{Dijkstra}. This work was supported by the
MRSEC Program of the National Science Foundation under Award
Number DMR-0820341.
\end{acknowledgments}


\begin{thebibliography}{26}
\bibitem{fn_ordermap}
This two-parameter description is but a very small subset of the
relevant parameters that are necessary to fully characterize a
configuration, but it nonetheless enables one to draw important
conclusions \cite{ToStRMP}. For example, another axes that could
be included is the mean contact number per particle, $Z$.
\bibitem{kuper_bound}
It is shown in Ref.~\cite{kuper2} that for any convex shape in
two-dimensions, the maximal packing density is bounded from below
by $\sqrt{3}/2 = 0.866025\ldots$.
\bibitem{fn_sc_packing}
This lattice packing corresponds to stacking layers
of aligned cylinders such as the semisphere caps are packed as spheres
in the face-centered cubic lattice.
\bibitem{fn_melting}
In future work, we will slowly compress dense liquid states of
truncated tetrahedra to examine whether the resulting crystalline
phases are the same as the ones that we identified via
decompression of the putative densest packing of this Archimedean
solid \cite{TrunTetrah}. Precise free energy computations will
provide of course the best way to ascertain the phase behavior of
truncated tetrahedra over the allowable range of densities.
\end{thebibliography}
\end{document}